\documentclass[12pt, a4paper]{article}

\usepackage[T1]{fontenc}
\usepackage[utf8]{inputenc}
\usepackage{color}
\usepackage{verbatim}
\usepackage{float}
\usepackage{amsmath}
\usepackage{mathrsfs}
\usepackage{amsfonts}
\usepackage{graphicx}
\usepackage{algpseudocode}
\usepackage{lineno,hyperref}
\newcommand{\ddim}{\textit{d}-dimensional }

\newcommand{\mc}{MC }

\makeatletter

\floatstyle{ruled}
\newfloat{algorithm}{tbp}{loa}
\providecommand{\algorithmname}{Algorithm}
\floatname{algorithm}{\protect\algorithmname}

\makeatother

\usepackage[english]{babel}

\begin{document}

\begin{flushleft}
{\large
\textbf{Neural Networks with Non-Uniform Embedding and Explicit Validation Phase to Assess Granger Causality}
}
\\
\vspace{5pt}
{\small
Alessandro Montalto$^{1,\ast}$, 
Sebastiano Stramaglia$^{2}$,
Luca Faes$^{3}$,
Giovanni Tessitore$^{4}$,
Roberto Prevete$^{5}$,
Daniele Marinazzo$^{1}$
}
\\
\vspace{5pt}
{\scriptsize
\bf{1} Data Analysis Department, Ghent University, Ghent, Belgium
\\
\bf{2} Dipartimento Interateneo di Fisica, University of Bari, and INFN Sezione di Bari, Italy
\\
\bf{3} BIOtech, Dept. of Industrial Engineering, University of Trento, and IRCS-PAT FBK, Trento, Italy
\\
\bf{4} Department of Physical Sciences, University of Naples Federico II
\\
\bf{5} DIETI, University of Naples Federico II
\\
$\ast$ E-mail: alessandro.montalto@ugent.be
}
\end{flushleft}



\begin{abstract}
A challenging problem when studying a dynamical system is to find the interdependencies among its individual components. Several algorithms have been proposed to detect directed dynamical influences between time series. Two of the most used approaches are a model-free one (transfer entropy) and a model-based one (Granger causality). Several pitfalls are related to the presence or absence of assumptions in modeling the relevant features of  the data. We tried to overcome those pitfalls using a neural network approach in which a model is built without any a priori assumptions. In this sense this method can be seen as a bridge between model-free and model-based approaches. The experiments performed will show that the method presented in this work can detect the correct dynamical information flows occurring in a system of time series. Additionally  we adopt a non-uniform embedding framework according to which only the past states that actually help the prediction are entered into the model, improving the prediction and avoiding the risk of overfitting. This method also leads to a further improvement with respect to traditional Granger causality approaches when redundant variables (i.e. variables sharing the same information about the future of the system) are involved.  Neural networks are also able to recognize dynamics in data sets completely different from the ones used during the training phase. 
\let\thefootnote\relax\footnote{Article published in Neural Networks
\\
http://www.sciencedirect.com/science/article/pii/S0893608015001574
\\
doi: http://dx.doi.org/10.1016/j.neunet.2015.08.003}

\end{abstract}




\section{Introduction}

A fundamental problem in the study of dynamical systems is how to find the
interdependencies among their individual components, whose activity is recorded and stored in time series. Over the last few years, considerable effort has been dedicated to the development of
algorithms for the inference of causal relationships among
subsystems, a problem which is strictly related to the estimate of
the information flow among subsystems \cite{wibral2014directed,sameshima2014methods}.  Two major
approaches to accomplish this task  are Granger causality (GC)
\cite{Granger_1969,bressler2011wiener}  and transfer entropy (TE)
\cite{schreiber2000measuring}. GC is based on regression, testing whether a source variable
(driver) is helpful to improve the prediction of a destination
variable (target) beyond the degree to which the target predicts its
own future. GC is a model-based approach, implying that the corresponding statistics for validation can be derived from analytic models, resulting in a fast and accurate analysis. 
A pitfall, however, is inherent to model-based approaches: the model assumed to explain the data often implies strong assumptions and the
method is not able to detect the correct directed dynamical networks
when these assumptions are not met. On the other hand non-parametric
approaches, such as transfer entropy, allow the pattern of influences to be obtained in the absence of any guidance or
constraints from theory; the main disadvantages of non-parametric
methods are the unavailability of analytic formulas to evaluate
the significance of the transfer entropy and the computational
burden, typically heavier than those required by model-based
approaches.

Feed-forward neural networks, consisting of layers of
interconnected {\it artificial neurons} \cite{Rumelhart:1986:PDP:104279}, are among the
most widely used statistical tools for non-parametric regression. Relying on neural networks, the proposed approach to Granger causality will be both non-parametric and based on regression, thus realizing the Granger paradigm in a non-parametric fashion.

In this paper we address the implementation of Granger's original
definition of causality in the context of the artificial neural networks
approach \cite{bishop1995neural}. The metrics used to validate the hypothesis of directed influence is the \textit{prediction error}: the difference between the network output and the expected target. The choice of the correct prediction error, and consequently the choice of the past states of the time series that will be fed to the model, has to be accompanied by a validation phase. Only under optimization of the \textit{generalization error} one can be sure that the network is not overfitting.  

In order to deal with an increasing number of inputs, each one representing a specific candidate source of directed influence, we will adopt a non-uniform embedding procedure \cite{PhysRevE.83.051112} that is an iterative procedure to select only the most informative past states of the system to predict the future of the target series among a wider number of available past states. In line with this procedure the network will be trained with an increasing number of inputs, each of them representing a precise past state of the variables that are most helpful to predict the target. Also this selection process will be implemented using the notions of prediction error and generalization error, the former quantifying how well the training data are reproduced, the latter describing the goodness of the validation on a novel set of data.

It is worth stressing that a neural networks approach to GC has been already proposed in \cite{attanasio2011detecting}, where  neural networks with a fixed number of inputs, together with other estimators of information flow, are used to evaluate GC. In \cite{attanasio2011detecting} neural  networks are trained without a validation set and an empirical method to avoid overfitting is adopted. To our knowledge the present approach is the first time that non-uniform embedding and a regularization strategy by a validation set are used together in the context of neural network approaches to detect dynamic causal links.
Moreover, the neural networks  built by our approach will accomplish not only the task of estimating information flows among variables, 
 they may also be used for dynamic classification task as well, as better explained in Subsection \ref{CT}. The new method presented in this work has been integrated in MuTE MATLAB toolbox\footnote{MuTE is a freeware toolbox. A detailed explanation is available on www.mutetoolbox.guru} \cite{montalto2014mute} and it will be compared here with the linear Granger causality as well as with the Transfer Entropy, both implemented in the non-uniform embedding framework.

\section{Introduction to neural networks}
\label{inn}

Artificial neural networks (ANN) are a very popular branch of machine learning. Here we give a brief introduction to neural networks to make this work self-consistent.

Neural networks can be represented as oriented graphs whose nodes  are simple processing elements called ``neurons'' handling their local input, consisting of a weighted summation of the outputs from the parents nodes \cite{Rumelhart:1986:PDP:104279}. The input signal is processed by means of a function, called ``activation function'', and the corresponding outcome, called ``output'', is then sent to the linked nodes  by a weighted connection;  the weight is a real number that represents the degree of relevance of that connection inside the neural network. The most common architecture of a neural network consists of neurons ordered into layers. The first one is called ``input layer'' that receives the external inputs. The last layer is called ``output layer'' that gives the result of the computations made by the whole network. All the layers between the input and output layer are called ``hidden layers''. 

Neural networks with at least one hidden layer and activation functions as the sigmoid function on the hidden nodes  are able to adequately approximate all kinds of continuous functions defined on a compact set from a \textit{d}-dimensional input space $\mathbb{R}^d$, the domain, to a \textit{c}-dimensional output space $\mathbb{R}^c$, the codomain given a sufficient number of hidden nodes: in this sense one can say that neural networks can perform any mapping between two different vector spaces \cite{bishop1995neural}. In order to allow a neural network to find the correct mapping, a so-called ``learning phase'' is needed. In this work we use \textit{supervised} learning, during which inputs are presented to the network and its output is compared to a known output. The weights are adjusted by the network that tries to minimize a cost function that depends upon the network output and the known output. This kind of learning allows a network to discover hidden patterns inside the data.

In this work we implement a growing neural network to study dynamical interactions in a system made up of several variables, described by time series, interacting with each other. The aim of the work is not only to find a directional relationship of influence between a subset of time series, the \textit{source}, and a \textit{target} time series taking into account the rest of the series collected in a set, called \textit{conditioning}, but also to determine the delay at which the source variables are influencing the target. We will then see how the neural networks approach can be useful to accomplish, under the same framework, several tasks such as: finding the directed dynamical influences among variables chosen at a certain delay; predicting a target series when the network is fed with a novel realization of a dynamical system whose connectivity structure has been previously learned; classifying a new data set, giving information about how close the causal relationships are to those observed in data sets used during the learning phase.

\subsection{Mathematical framework}
\label{mathFrame}

In this work we deal with growing feed-forward neural networks to better infer the directed dynamical influences in a composite system. Each stochastic variable at hand is assumed to be zero mean (the mean of the data sample is subtracted from data), hence we will deal with neural networks without bias terms. A classical feed-forward neural network without bias is usually described by: a finite set of \textit{O} nodes $S={1,2,\ldots, O}$ divided in \textit{d} inputs, \textit{c} output nodes and $O-(d+c)$ hidden nodes; a finite set of one way direction connections \textit{C} each one connecting a node belonging to the $\mathsf{k}$-th layer to a node belonging to the $\mathsf{h}$-th layer, with $\mathsf{h} > \mathsf{k}$. A weight $w_{hk}$ is associated with each connection from the node $k \in \mathsf{k}$ to the node $h \in \mathsf{h}$. Each node \textit{o} is characterized by an input function $s_o$, an input value $i_o$, 
an activation function $f_o$, and an output value $z_o$ \cite{bishop1995neural}.
 Let us now define $\mathbf{w}_h$ as the weights vector of the connections which leave the nodes of 
the $\mathsf{k}$-th layer and reach the node $h$. Let us define $\mathbf{z}$ as the output vector of a generic layer of nodes. The input $i_h$ is given by $i_h = s_h(\mathbf{w}_h, \mathbf{z})$. Usually we have: $i_h = \sum_k{w_{hk} \cdot z_k}$. 
Each $z_h$ is given by 

\begin{equation}
z_h  = f_h\left(\sum_k{w_{hk} \cdot z_k}\right)
\label{eq:nodeoutput}
\end{equation}

\begin{figure}[ht!]
	\centering
    \includegraphics[width=0.8\textwidth]{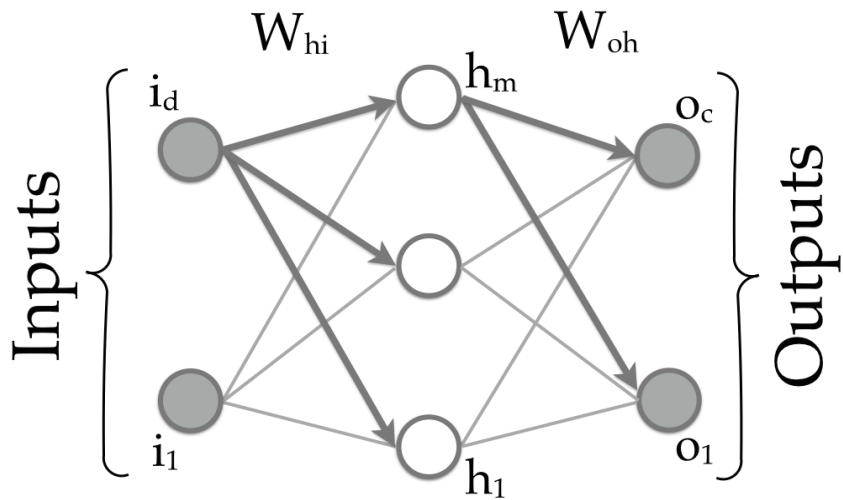}
    \caption{Schematic representation of a feed-forward neural network.}
     \label{neunetscheme}	
 \end{figure}

To evaluate the output of a multilayer network, consecutive applications of \eqref{eq:nodeoutput} are needed to activate all network nodes. Figure \ref{neunetscheme} depicts the schematic structure of the feed-forward networks under discussion here.

To summarize: the output values of a network can be expressed as deterministic functions of the inputs. Assuming that the network has only one hidden layer $\mathsf{h}$, we can say that the whole network represents a function, linear or non-linear depending on the linearity or non-linearity of the $f_h$, between the \ddim input space and the \textit{c}-dimensional output space, with parameters \textbf{w} given by the network weights. The relation holding between inputs and outputs of the network can be approximated in the multidimensional space spanned by the hidden nodes by either an hyperplane if linear functions are used as activation functions of the hidden nodes or by a smoother approximation when non-linear functions are set as activation functions of the hidden nodes. Usually, the activation functions of the output nodes are set to be the identity function that does not modify the input values of the output nodes because the outputs of the network are not supposed to be bounded in order to assume values as close as possible to the training target values, assuming that overfitting is avoided.

So far we have shown how neural networks can process inputs and how they can be mapped onto a parametric function $\mathcal{F}(\mathbf{x};\mathbf{w}) :
\mathbb{R}^d \rightarrow \mathbb{R}^c$.

We can now assume that there is a function $f : \textbf{x} \in \mathbb{R}^d \rightarrow f(\textbf{x}) \in \mathbb{R}^c$ to be modelled and we know a finite set of \textit{N} couples $(\textbf{x}^n, \textbf{t}^n)$, where $n \in [1,\ldots , N]$, $\textbf{t}^n$ is the value of the function $f(\textbf{x})$ evaluated in $\textbf{x}^n$ plus an error $\epsilon(\textbf{x}^n)$. We want to approximate \textit{f} using the parametric function $\mathcal{F} : \mathbf{w} \in \mathbb{R}^p, \mathbf{x} \in \mathbb{R}^d \rightarrow \mathcal{F}(\mathbf{x};\mathbf{w}) \in \mathbb{R}^c$. The function $\mathcal{F}$ can be found through the minimization of a certain error function $E(\mathbf{w})$. For instance a classical error function is the sum of squares function \eqref{eq:error} to minimize by means of an iterative procedure that requires the data to be presented to the network several times through consecutive realizations called \textit{epochs}:

\begin{equation}
	E = \frac{1}{2} \sum^{N}_{n=1}\sum^d_{k=1}{(y^n_k - x^n_k)^2}
	\label{eq:error}
\end{equation}

The training of the network is the process to determine, starting from a finite set of couples $(\mathbf{x}^n, \mathbf{t}^n)$, the weights $\tilde{\mathbf{w}}$ that can better shape $\mathcal{F}$ to be as close as possible to \textit{f}. After each epoch of the training phase the weights in the network are adjusted. At this point a definition of \textit{close} is in order. Let us suppose a noisy dataset consisting of $\mathbf{x}^n$ and $\mathbf{t}^n = f(\mathbf{x}^n)+ \epsilon(\mathbf{x}^n)$ where $\epsilon$ is the noise term. If we train the network until the input can be exactly reproduced then $\mathcal{F}$ is not only reproducing \textit{f}, but the noise too. It is easy to understand that the more specialized the network the less it will be able to predict the right $\mathbf{t}^{n'} = f(\mathbf{x}^{n'})+ \epsilon(\mathbf{x}^{n'})$ when a $\mathbf{x}^{n'}$ never seen before is presented to the network. In this case we say that the network is not able to \textit{generalize}. To overcome this issue the \emph{validation phase} is embedded in the learning. The validation phase is paramount because it allows the network to both model the function from which the data could have been drawn and to avoid modeling fluctuations produced by noise in the training set. In order to accomplish these two modeling tasks at the same time, the whole learning procedure is divided into two well distinguished steps:

\begin{enumerate}
 \item the whole data set is divided in two groups. One group is used for the training step during which the weights are updated
 \item the second group is used for the validation step. These data have not been used in the previous step. We used a maximum amount of {\it training epochs} and a smaller number of epochs called \textit{validation epochs}. The validation phase is embedded in the learning phase: this combination of training and validation avoids erroneous use of the training procedure, thus avoiding overfitting.
\end{enumerate}

In the following section, we present the algorithm used for the learning phase:

\begin{enumerate}
 \item training step: adjust the weights after a number of {\it training epochs}
 \item validation step: evaluate the generalization error and store it in a vector \textit{VEvect}
 \item repeat steps 1. and 2. continuing to train the network until one of the following three stop conditions is verified: 
 \begin{itemize} 
\item the relative error evaluated as 
$$\frac{\Vert \mbox{current VEvect entry} - mean(\mbox{previous VEvect entries})\Vert}{\mbox{current VEvect entry}}$$
is less than a \textit{validation threshold} set to $10^{-3}$. The value of \textit{previous VEvect} is set to 5; 
\item 
$$\frac{\left(\mbox{current VEvect entry} - mean\left(\mbox{previous VEvect entries}\right)\right)}{\mbox{current VEvect entry}} \geq 0$$
\item the maximum number of {\it training epochs} is reached. \label{plateau}
\end{itemize}
\end{enumerate}
The \textit{previous VEvect} and \textit{validation threshold} values have been chosen taking into account a cautious gradient descent implying small updating steps as the main concern here is the risk of overfitting.

\section{Granger Causality with neural networks}
\label{causality}

The aim of this work is to find directed dynamical influences among variables, modeled as time series, using neural networks as a powerful tool to compute the prediction errors needed to evaluate causality in the Granger sense. According to the original definition, Granger causality (GC) deals with two linear models of the present state of a target variable. The first model does not include information about the past states of a driver variable, while the second model contains such information. If the second model's error is less than that of the first model in predicting the present state of the target, then we can safely say that the driver is causing the target in the sense of Granger \cite{granger1969investigating}. 
Here we introduce a new Granger causality measure called \textit{neural networks Granger causality} (NNGC) defined as

\begin{equation}
NNGC = err_{\mbox{reduced}} - err_{\mbox{full}}
\label{nnc}
\end{equation}
where $err_{\mbox{reduced}}$ is the prediction error obtained by the network that does not take into account the driver's past states, while $err_{\mbox{full}}$ is the prediction error evaluated by the network that takes into account the driver's past states.

Therefore, instead of fitting predefined models, (linear ones in the original proposal by Granger) we train a neural network to estimate  the target using only the past states that can better explain the target series, by using the non-uniform embedding technique. Such strategy leads to growing neural networks,  with an increasing number of input neurons, each input neuron representing a past state chosen from the amount of past states available, considering all the variables in the system. The architecture of the network and choice of the most suitable past states, performed through the non-uniform embedding approach, are described in detail in the next sections. Relying on
neural networks, this method realizes the Granger paradigm in a non-parametric fashion, like in \cite{PhysRevE.70.056221,PhysRevE.73.066216} where radial basis function networks where employed. This article improves such previous work by (i) using non-uniform embedding and (ii) employing training and validation phases concurrently to ensure a more robust detection of dynamical interactions.

\section{Non-uniform embedding (NUE)}
\label{nue}

We first introduce in this section the NUE approach which is the basis of the algorithm used to build a neural network able to find the correct mapping between the input and the output spaces in an optimal way. The uniform embedding (UE) approach relies on a predefined set of candidates making a strong assumption about which past states are better able to explain the future of the target series. This approach, lacking a specific criterion according to which the candidates are chosen, is likely to cause problems such as overfitting and detection of false influences \cite{PhysRevE.75.056211,marinazzo2008kernel}. NUE framework, instead, is an iterative procedure aimed at detecting only the time series' past states that can effectively help to predict the target series. To evaluate whether a new candidate should be chosen, an hypothesis and, eventually, a significance test, should be satisfied. In this way of exploring causality, once this hypothesis and significance test (when needed) are no longer satisfied, the procedure is unable to find additional candidates to help predict the target.

Let us consider a composite system described by a set of $M$ interacting dynamical
(sub) systems and suppose that, within the composite system, we are interested in evaluating the information flow from the source system $\mathcal X$ to the destination system $\mathcal Y$, collecting the remaining systems in the vector $\textbf{$\mathcal Z$} = \left\{\mathcal Z^k\right\}_{k = 1,\ldots,M-2}$. We develop our framework under the assumption of stationarity, which allows to perform estimations replacing ensemble averages with time averages (for non-stationary formulations see, e.g., \cite{ledberg2012framework} and references therein). Accordingly, we denote \textit{X}, \textit{Y} and \textbf{Z} as the stationary stochastic processes describing the state visited by the systems $\mathcal{X}$, $\mathcal{Y}$ and $\mathcal{Z}$ over time,
and $X_n$, $Y_n$ and $\textbf{Z}_n$ as the stochastic variables obtained sampling the processes at the present time \textit{n}. Moreover, we denote $X_n^-=[X_{n-1}X_{n-2}\ldots]$, $Y_n^-=[Y_{n-1}Y_{n-2}\ldots]$, and $\textbf{Z}_n^-=[\textbf{Z}_{n-1}\textbf{Z}_{n-2}\ldots]$ as the infinite-dimensional vector variables representing the whole past of the processes \textit{X}, \textit{Y} and \textbf{Z}. In some cases, taking the instantaneous influences of the candidate drivers into account as well may also be desirable. In such cases, the vectors $X_n^-$ and $\textbf{Z}_n^-$ defined above should also contain the present terms $X_n$ and $\textbf{Z}_n$.

We will discuss here the crucial issue of how to approximate the infinite-dimensional variables representing the past of the processes. This problem can be seen in terms of performing suitable conditioned embedding of the considered set of time series \cite{vlachos2010nonuniform}.

The main idea is to reconstruct the past of the whole system represented by the processes $X, Y, \textbf{Z}$ with reference to the present of the destination process $Y$, in order to obtain a vector $V = [V_n^Y, V_n^X, V_n^{\textbf{Z}}]$ containing the most significant past variables to explain the present of the destination. 

Non-uniform embedding constitutes the methodological advance with respect to the state of the art that we propose as a convenient alternative to UE. This approach is based on the progressive selection, from a set of candidate variables including the past of $X$, $Y$, and \textbf{Z} considered up to a maximum lag (\textit{candidate set}), of the lagged variables which are more informative for the target variable $Y_n$. At each step, selection is performed maximizing the amount of information that can be explained about $Y$ by observing the variables considered with their specific lag up to the current step. This results in a criterion for maximum relevance and minimum redundancy for candidate selection, so that the resulting embedding vector $V=[V_n^X\, V_n^Y\, V_n^Z]$ includes only the components of $X_n^-$, $Y_n^-$ and $\textbf{Z}_n^-$, which contribute most to the description of $Y_n$. Starting from the full candidate set, the procedure 
which prunes the less informative terms is described below:

\begin{enumerate}
 \item Get the matrix with all the candidate terms \\ \mc $=[X_{n-1} \ldots X_{n-l_X} Y_{n-1} \ldots Y_{n-l_Y} \textbf{Z}_{n-1} \ldots \textbf{Z}_{n-l_Z}]$, with $l_X$, $l_Y$, $l_Z$ representing the maximum lag considered for the past variables of the observed processes; these matrices will contain also the terms $X_n$ and $\textbf{Z}_n$ in case one wants to take into account instantaneous effects. The values of $l_X$, $l_Y$, $l_Z$ can be set by the experimenter according to a known feature of the data, or set to a reasonably large value for exploratory purposes. If values of $l_X$, $l_Y$ and $l_Z$ are set too low, an incorrect estimation of Granger causality may result, but higher values should not issues with non-uniform embedding.
 \item Run the procedure to select the most informative past variables and the optimal embedding vector:
 	\begin{enumerate}
 	\item Initialize an empty embedding vector $V_n^{(0)}$
        \item Perform a while loop on $k$, where $k$ can assume values from 1 to the number of initial available candidates, \textit{numC}, in the \mc matrix. At the $k-$th iteration, after having chosen $k-1$ candidates collected in the vector $V_n^{(k-1)}$:\\
     for $1 \leq i \leq $ number of current candidate terms \\
		\begin{itemize}
         \item  add the $i-$th term of \mc, $W_n^{(i)}$, to a copy of $V_n^{(k-1)}$ to form the temporary storage variable $V_n' = [W_n^{(i)}V_n^{(k-1)}]$\\
			\item compute the information exchanged between $Y_n$ and $V_n'$\\
			\end{itemize}
			
     \item Among the tested $W_n^{(i)}$, select the term $\hat{W}_n$ which maximizes the information exchanged \\ 
     \item \textbf{if} $\hat{W}_n$ satisfies a \textit{termination criterion},  delete it from \mc and set $k = k + 1$. \label{ifcond} \\		
     \item \textbf{else} end the procedure setting $k=numC+1$ and  returning $V=V_n^{(k-1)}$
 	\end{enumerate}
 \item Use $Y_n$ and the full embedding vector $V=[V_n^X\, V_n^Y\, V_n^Z]$ and to evaluate $err_{\mbox{full}}$. $err_{\mbox{reduced}}$ is obtained excluding from $err_{\mbox{full}}$ the candidates belonging to the variables considered as drivers. Both errors are evaluated as the root mean squared error (RMSE) between the neural network output and $Y_n$. If the error resulting from the network that contains the inputs representing the driver's past states ($err_{\mbox{full}}$) is lower than the error resulting from the network that does not take into account the driver's past states ($err_{\mbox{reduced}}$), then the driver assessed is determined to help predict the target more than the network that excludes the driver.
\end{enumerate}

The complexity of the algorithm concerns mainly step 2, in particular step 2(b), involving a \textit{for} loop nested inside a \textit{while} loop: in the worst case the body of the \textit{for} loop is executed $numC^2$ times resulting in a complexity $\mathcal{O}(numC^2)$.

Summarizing, the non-uniform embedding is a feature selection technique selecting, among the available variables describing the past of the observed processes, the most significant - in the sense of predictive information - for the target variable. Moreover, given the fact that the variables are included into the embedding vector only if associated with a significant contribution to the description of the target, the significance of the NNGC estimated with the NUE approach results simply from the selection of at least one lagged component of the source process. In other words, if at least one component from X is selected by NUE, the estimated NNGC is strictly positive and can be assumed to be significant. If not, the estimated NNGC is exactly zero and is assumed to be non-significant. This latter also occurs when the first candidate ($k=1$) does not reach the desired level of significance, meaning that none of the candidates provides significant information about the target variable. This may also be encountered, for instance, when the target process consists of white noise: the code will return an empty embedding vector and assign a zero value to the NNGC.

\section{Non-uniform embedding using neural networks (NeuNet NUE)}
\label{sec:neunetnue}

Here we want to investigate the opportunity to use neural networks to create the two models needed to evaluate NNGC and, at the same time, to better choose the right candidates to be considered as terms of the models. In this sense our method is a model-free approach because we do not assume any model a priori that can explain the data, but we allow the network to explore the parameters space in order to find the model we need. The procedure will be able to model a function from the input space, spanned by the time series' past states, and the output space, spanned by the present state of the target series: $Y_n = f(V)$. It will be possible to estimate a function $\mathcal{F}$ as close as possible to $f$. This will ensure a precise prediction of \textit{Y} from \textit{Y} itself, \textit{X} and \textbf{Z}. It is easy to see that from $\mathcal{F}$ it is possible to assess whether for another data set $Y',X', \mathbf{Z}'$, the same relation $\mathcal{F}$ holds: 
in this case the network will be able to \textit{generalize}.

In this study a three-layers feed-forward neural network is used \cite{bishop1995neural}, trained by means of the resilient back propagation technique that is one of the fastest learning algorithms \cite{riedmiller1993direct}. Briefly, the resilient back propagation is an optimized algorithm to update the weights of a neural network based on the gradient descent technique. Let $\Delta_{ij}$ be the weight update value that only determines the size of the weight update and \textit{E} the error function. Then the resilient back propagation rule is the following:

\begin{equation}
\Delta_{ij} = \left\{
  \begin{array}{l l}
    \eta^+ \times \Delta^{(t-1)}_{ij} & \quad \text{if $\frac{\partial E}{\partial w_{ij}}^{(t-1)} \times \frac{\partial E}{\partial w_{ij}}^{(t-1)} > 0$}\\
    \eta^- \times \Delta^{(t-1)}_{ij} & \quad \text{if $\frac{\partial E}{\partial w_{ij}}^{(t-1)} \times \frac{\partial E}{\partial w_{ij}}^{(t-1)} < 0$}\\
    \Delta^{(t-1)}_{ij} & \quad \text{else}
  \end{array} \right.
\end{equation}
where $0 < \eta^-<1<\eta^+$. To summarize: every time the partial derivative of the current weight $w_{ij}$ changes in sign, i.e. the error function slope changes indicating that a local minimum has been avoided, the updated value $\Delta_{ij}$ is decreased by the factor $\eta^-$ allowing a reversal, or ``coming back'', in the parameters space towards the local minimum. If the derivative does not change sign, then the updated value $\Delta_{ij}$ is increased by the factor $\eta^+$ accelerating towards a local minimum.

Once the updated value is evaluated, the weight update is quite straightforward as shown by the following equations:

\begin{equation}
\Delta^{(t)}_{w_{ij}} = \left\{
  \begin{array}{l l}
    -\Delta^{(t)}_{ij} & \quad \text{if $\frac{\partial E}{\partial w_{ij}}^{(t)}  > 0$}\\
    +\Delta^{(t)}_{ij} & \quad \text{if $\frac{\partial E}{\partial w_{ij}}^{(t)} < 0$}\\
    0 & \quad \text{else}
  \end{array} \right.
\end{equation}
so that $w^{(t+1)}_{ij} = w^{(t)}_{ij} + \Delta^{(t)}_{w_{ij}}$. However, we should also take into account the case when the partial derivative changes sign: the previous weight update is then reverted as follows:

\begin{equation}
\begin{array}{l l}
\Delta^{(t)}_{w_{ij}} = - \Delta^{(t-1)}_{w_{ij}} & \quad \text{if $\frac{\partial E}{\partial w_{ij}}^{(t-1)} \times \frac{\partial E}{\partial w_{ij}}^{(t)} < 0.$}
\end{array}
\end{equation}

Following the NUE scheme, each input corresponds to a candidate, while the minimization criterion is the prediction error between the network output and $Y_n$. We should keep in mind that the core of the entire procedure lies in the choice of the candidates that can actually help to predict the target series. Once the relative prediction error, defined as $(\mbox{prediction error}_{k-1}-\mbox{prediction error}_k)/(\mbox{prediction error}_1-\mbox{prediction error}_k)$ where \textit{k} can assume values from 1 to the number of initial available candidates, is greater than or equal to a threshold, the procedure stops and no further candidates are chosen. To summarize: the hypothesis of Granger causality evaluates how much information is introduced by adding a new input with respect to the information carried only by the inputs previously considered. Moreover, it is worth stressing that in this case we do not rely on the comparison with a null distribution in order to choose whether a given candidate must be chosen or not. On the other hand, when a driver-response relationship among variables holds, the algorithm will find, input by input, the candidate that will give the lowest prediction error, this being a condition that can hold only if we ensure the network is not overfitted. The risk of overfitting is the reason why a validation phase, described in detail in the following sections, was implemented and the idea of a fixed amount of training iterations was discarded. As soon as the error on the validation set, called \textit{generalization error} increases, the training of the network stops ensuring the capability of the network to generalize, implying that it has not been overfitted.

To better explain the steps implemented to select the past states as a pseudo-code we can say that a \textit{for} loop is nested within a \textit{while} loop. The \textit{while} loop condition, that takes into account the decreasing of the prediction error during the training phase, determines whether or not an additional input should be added to the network. It is worth stressing that during the whole procedure of the candidates' selection, the internal architecture of the network is kept fixed: the number of hidden nodes is set up as a fraction of the maximum number of candidates available, as shown in subsection \ref{correct_params}, and it does not change. The \textit{for} loop, instead, tests each available candidate given the previous inputs already chosen. During this test, at each iteration of the \textit{for} loop, a network is trained taking into account the current candidate and the validation phase takes place according to the procedure explained in Subsection \ref{mathFrame} point \ref{plateau}. Therefore, the validation error is taken into account in order to allow the network itself to reach its best performance, in terms of the generalization task, according to the current candidate. At the end of the \textit{for} loop the candidate which gives the minimum prediction error is selected. If the prediction error satisfies the \textit{while} loop condition, such that the relative prediction error is smaller than a threshold, the candidate is chosen and deleted from the set of the available candidates so that the procedure can continue. Otherwise, the procedure will stop. The pseudo-code of the algorithm is shown in the following:

\begin{footnotesize}
\begin{algorithmic}[1]
\State {Initialize network parameters;}
\State {Initialize the embedded matrix \textit{EM} $= \emptyset$}
\State {Initialize the prediction error \textit{PE} vector $= \emptyset$}
\State {Initialize the current prediction error \textit{CPE} vector $= \emptyset$}
\State {Initialize final candidate matrix FCM $= \emptyset$}
\State {Initialize CS $=[X_{n-1} \ldots X_{n-l_X}, Y_{n-1} \ldots Y_{n-l_Y}, \textbf{Z}_{n-1} \ldots \textbf{Z}_{n-l_Z}]$. The terms $X_n$ and $\textbf{Z}_n$ should be considered too in case the instantaneous effects should be taken into account.}
\State {k = 1}
\While {CS $\neq \emptyset$} \label{externalwhile}
  \If {CS is full}
    \State{Initialize the network NN with one input, the number of chosen hidden nodes, one output}
  \Else
    \State{Add to NN$_{k - 1}$ another input;}
    \State {Initialize only the weights between the new input and the hidden nodes keeping all the rest fixed;}
  \EndIf
   \For {$i \in [1, \ldots, \mbox{length(CS)}]$} \label{forloop}
     \While{epoch $\leq$ maxTrainEpochs} \label{internalwhile}
      \Statex {\% Learning phase:}
      \State {train the network, after 30 training epochs evaluate the prediction error;}
      \State {validate the network evaluating the generalized error;}
       \If {epochs/valEpochs == 0 \textbf{and} epochs $\leq$ maxTrainEpochs} \label{starteval_val}
           \State {evaluate the relative validation error}
           \If {$\Vert \mbox{relative validation error}\Vert \leq$ validationThreshold \textbf{or} relative error $\geq 0$ \textbf{or} epochs == maxTrainEpochs}
               \State {Store the prediction error in CPE(i)}
               \State{epoch = maxTrainEpochs + 1}
           \EndIf
           \State {Store the prediction error in CPE(i);}
           \State{epoch = epoch + 1}
       \EndIf \label{stopeval_val}
     \EndWhile
     \EndFor
     \State{NN$_k$ =  neural network having in input the candidates that give the minimum prediction error stored in CPE}
     \State{PE$_k$ = min(CPE)}
   \If {relative prediction error $\leq$ trainThreshold} \label{starteval_train}
      \State{NN = NN$_{k - 1}$}
      \State{PE = PE$_{k - 1}$}
      \State{CS $= \emptyset$}
   \Else
      \State{NN = NN$_{k}$}
      \State{add to FCM the candidate of CS that returns the minimum prediction error}
      \State{delete from CS the candidate that returns the minimum prediction error}
      \State{k = k + 1}
   \EndIf \label{stopeval_train}
 \EndWhile
 \State{return NN; FCM; PE}
\end{algorithmic}
\end{footnotesize}
where the strings after the \% symbol should be considered as non executable code.

In the following we will explain in more detail the algorithm showed above:
\begin{enumerate}
	\item In the initialization phase it is worth noting that $l_X$, $l_Y$, $l_Z$ represent the maximum lag considered for the past variables of the observed processes. In the following experiments we will set $l_X$, $l_Y$, $l_Z$ to take into account more past states than needed.
	\item at the \textit{k}-th step of the while loop at line \ref{externalwhile}, where \textit{k} runs on the number of inputs chosen, the network tests all the candidates available by means of the for loop at line \ref{forloop}: there are \textit{k} inputs. The first \textit{k}-1 inputs are the ones chosen so far and on the \textit{k}-th input one candidate per time is considered. The initial conditions are the same for each candidate: the weights have been fixed so the ones departing from the \textit{k}-1 inputs are the same found as the result of the training at the (\textit{k}-1)-th step and the weights departing from the \textit{k}-th input are the same at the beginning of each training session when the RMSE between the network output and $Y_n$ is evaluated. Lines \ref{starteval_val}-\ref{stopeval_val} take care of whether to stop the training phase for the current candidate according to the generalization error. \label{start}
	\item lines \ref{starteval_train}-\ref{stopeval_train} check whether it would be worth adding candidates, or it is better to stop the whole procedure because no further meaningful information can be added to better predict the target. \label{stopCondition}
	The generalization error is not relevant at this stage, since it is only used to stop the training phase.
	\end{enumerate}
The network is finally trained to reproduce the best correspondence between the space spanned by the terms of FCM and the space spanned by $Y_n$. The network is then the model that can be used to explain $Y_n$, including the driver's candidates. This model will give $err_{\mbox{full}}$. To evaluate $err_{\mbox{reduced}}$ the candidates belonging to the source system should be removed from the network, so the corresponding inputs and weights should not be considered. This configuration leaves the weights between the hidden nodes and the output unchanged, so the network now won't be able to approximate $Y_n$ as well as during the evaluation of the previous term if the causal hypothesis holds between the driver and the target. $err_{\mbox{reduced}}$ can be computed projecting the information carried by the inputs representing the candidates belonging to the target and the conditioning variables on the output space and evaluating the RMSE between the network output and $Y_n$. NNGC is now immediately evaluated by the difference between the two terms. The significance of the causality measure estimated with the neural network method embedded into the NUE approach results simply from the selection of, at least, one lagged component of the driver. In other words, if at least one component from the driver is selected, the Granger causality is strictly positive and can be assumed as significant. If this is not the case, the estimated causality that results is exactly zero and is assumed to be non-significant.

NUE is used here as a feature selection algorithm. Other feature selection algorithms can be used to select the most informative candidates; in the present work our choice is in line with other approaches to detect dynamical interactions present in literature, thus offering a coherent framework for all the estimators.

\section{Applications to simulated data}

Before applying the proposed method, the correct initialization parameters were set (see Subsection \ref{correct_params}). First of all, we wanted to prove that neural networks implemented with the non-uniform embedding framework perform better than neural networks implemented with the uniform embedding framework, Subsection \ref{discardue}. Then, several datasets were simulated to test NeuNetNUE in different situations in order to explore its capability to detect the correct directed dynamical links, see Subsections \ref{simHM} - \ref{simLS}. During those three experiments we compared the neural networks with a model-based approach and two model-free approaches, as described in Subsection \ref{simHM}, to get a better idea of the performances obtained by our method. Furthermore, we wanted to check whether NeuNetNUE was both robust with respect to redundant information (see Subsection \ref{simRD}) and able to outperform an approach based on multivariate Granger causality analysis \cite{stramaglia2014synergy}. Finally we wanted to evaluate the capability of the networks to predict and classify time series (see Subsection \ref{CT}). 

\subsection{Choice of the parameters}
\label{correct_params}

One of the crucial aspects of neural networks approaches concerns the choice of the optimal parameters. In this paper we are interested not only in the parameters involving the architecture and the training of the network, but also in the parameters that are responsible for the number of past states that can be chosen, allowing the approach to be more or less conservative. The parameters can be listed as follows:

\begin{itemize}
	\item the threshold according to which a certain number of past states are chosen (th). This parameter is taken into account to stop the training of the network and it consequently regulates the amount of past states chosen by the network: for a lower \textit{th} more past states are selected, see Section \ref{sec:neunetnue} pointed list \ref{stopCondition}
	\item the validation threshold, useful to not overfit the network (valTh). This parameter plays an important role in the validation phase, allowing the network not to be overfitted, see Section \ref{sec:neunetnue} pointed list \ref{start}
	\item the number of hidden nodes (hidNodes), reported as percentage of the total amount of the available past states
	\item the learning rates for the resilient back propagation, $\eta^+ , \eta^-$.
\end{itemize}

The parameters mentioned above must be set so that the neural network approach is able to detect the expected information flow. The investigation of the best parameters values was performed on linear and non-linear models with memory up to 3 points in the past.

%
We considered 20 simulations of the systems for each combination of the parameters values shown in table \ref{tab1}. We set $l_X = l_Y = l_Z = 5$. Values of $\eta^-$ and $\eta^+$, the parameters of the resilient back propagation, ranged as $\eta^- \in [0.4,\ldots,0.9]$ with step of 0.1 and $\eta^+ \in [1.1,\ldots,1.4]$ with step of 0.1. The threshold's values for the prediction error range between $2^{-8}$ and 0.3. According to this assumption, it is then possible to consider whether the current candidate is significant or not. Small values of the threshold, such as $2^{-8}$, represent a weakly conservative network. On the other hand, high values of the threshold, such as 0.3, represent a strongly conservative network. We first investigated how the network performed for higher values of the threshold and we found that the networks were too conservative and, consequently, NueNet NUE only found one candidate belonging to the target series only. The same reasoning holds for the validation threshold that gives the range of values within which the validation error can fall. The assumption on how wide the range is, determines whether the network can be considered to have undergone enough training. Finally the number of hidden nodes ranges between 0.1 and 2.5, with step of 0.2, times the number of available past states. In this way, we allow the network to have a number of hidden nodes so as to allow it to reach the best performance with increasing number of inputs. It turned out that \textit{number of hidden nodes} = 1.3$\times$ \textit{number of available candidates} was the best compromise.

\begin{figure}[ht!]
	\centering
    \includegraphics[width=1\textwidth]{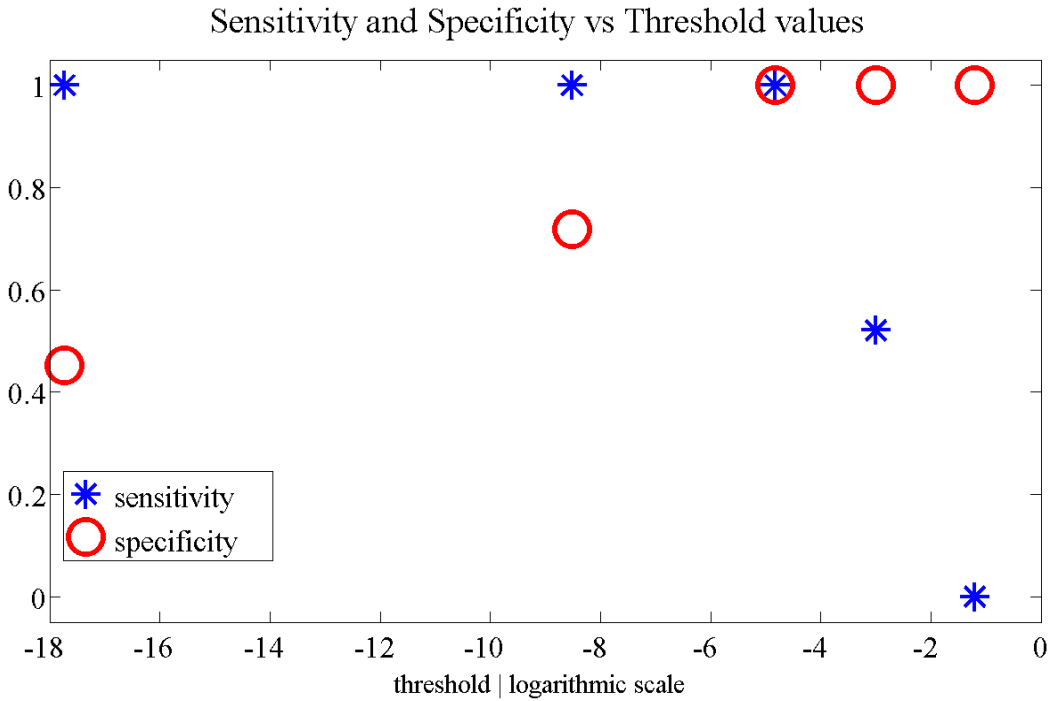}
    \caption{Sensitivity and specificity at varying of the threshold values.}
     \label{sensspec_vs_th}	
 \end{figure}

For each combination of the parameters we evaluated how many times the method was able to detect the right information flows, estimating the number of true positives (TP), true negatives (TN), false positives (FP)  and false negatives (FN). We then evaluated sensitivity = TP /(TP+FN), specificity = TN / (TN+FP) and $\mbox{F1}_{\mbox{score}}$ = 2\,TP/(2\,TP + FP + FN) and we checked how the performances changed as the parameters varied. We found that neural networks obtained high performances on both systems corresponding to different parameters values: parameters values obtained on the linear system allowed NeuNet NUE to be less conservative with regard to neural networks used with parameters values found on the non-linear system. We finally chose the parameters in correspondence to which the network would be considered less conservative: th = $8^{-3}$; valTh = 0.6; hidNodes = 0.3. Figure \ref{sensspec_vs_th} shows the performances of the network for different values of th, keeping all the other parameters fixed. For a better visualization, only five points out of the nine showed in the table were plotted. The other four points have been omitted, being too close to the others in the figure: the resulting curve is virtually unchanged. We can notice that for the minimum value of th, $2^{-8}$, the network  takes into account more past states than needed retrieving more FP and less TP than for higher values of th. Furthermore, the specificity is lower, almost zero, than the sensitivity. For th = $8^{-3}$ the network's performances give sensitivity = specificity = 1, while for the maximum value of th, 0.3, the network is not allowed to choose a lot of past states and, consequently, there are less TP and more TN than for lower values of th, resulting in the sensitivity lower than the specificity.

\begin{table}[ht]
	\centering
	\vspace{2mm}
		\begin{tabular}{cc}
		   \textbf{Name Parameter} & \textbf{Parameter values}\\
			\hline
			th & $2^{-8}\; 2^{-6}\; 2^{-4}\; 2^{-3}\;  5^{-3}\;  8^{-3}\;  1^{-2}\;    0.15 \; 0.30$\\
			\hline
			valTh & 0.2 0.4 0.6 0.8 1\\
			\hline
			hidNodes & 0.1 0.3 0.5 0.7 0.9 1.1 1.3 1.5 1.7 1.9 2.1 2.3 2.5\\
			\hline
			$\eta^-$ & 0.4 0.5 0.6 0.7 0.8 0.9\\
			\hline
			$\eta^+$ & 1.1 1.2 1.3 1.4\\
		\end{tabular}
		\caption{Parameters values to initialize the network.}
		\label{tab1}
	\end{table}
	
\subsection{Reasoning behind our discarding of the uniform embedding approach}
\label{discardue}
 
 Before explaining the experiments that we performed to test the proposed approach, we would like to underline that the non-uniform embedding framework was chosen because of its theoretical advantages with respect to the uniform embedding. Furthermore, we wanted to check whether those advantages held in the case of the GC neural networks estimator too. Neural networks used with the uniform embedding approach (NeuNet UE) only need one network to be trained when $err_{\mbox{full}}$ is evaluated. Each input of the network represents a past state, therefore the number of inputs equals the number of available past states. The other network parameters have the same values as in the case of NeuNet NUE. The validation phase is still required. Once $err_{\mbox{reduced}}$ is evaluated by only removing the inputs corresponding to the past states that belong to the driver whose influence to a specific target is tested, we can obtain a value of NNGC whose significance still has to be evaluated. This step is addressed using surrogates technique as implemented in the case of other estimators also used into the uniform embedding framework \cite{montalto2014mute}. This means that for each surrogate another network with the same architecture has to be trained resulting in a dramatic increase of the computational complexity.
 
\begin{figure}[ht!]
\centering
    \includegraphics[width=0.7\textwidth]{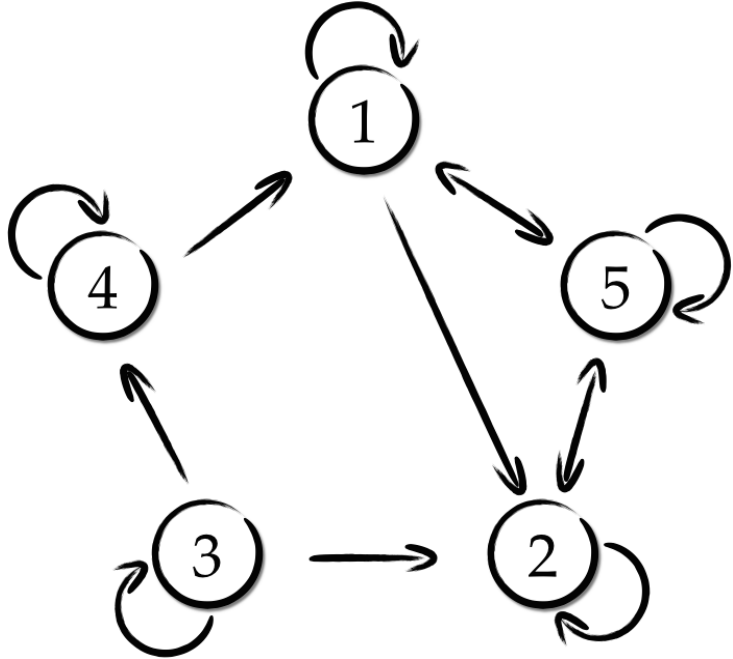}
    \caption{Simulated system. Interactions between the variables of the simulated H\'enon maps system generated according to equations \eqref{eq:henonAle}.}
    \label{simsysHenonAle}
\end{figure}

We compared NeuNet NUE and NeuNet UE performing a multivariate analysis on 100 realizations of a system composed of five coupled H\'enon maps with a length of 2500 time points, built according to the following equations:
 
 \begin{equation}
 \begin{aligned}
    X_{1,n} &= aV(1) - (0.5c(X_{4,t-1}+X_{5,t-1}) + \\
               &  (1-c)X_{1,t-1})^2 + aV(2)X_{1,t-2} + w_{1,n}\\
    X_{2,t} &= aV(1) - (0.5c(X_{3,t-1}+X_{5,t-1}) + \\
               & (1-c)X_{1,t-1})^2 + aV(2)X_{1,t-2} + w_{2,n}\\
    X_{3,t} &= aV(1) - X_{3,t-1}^2 + aV(2)X_{3,t-2} + w_{3,n}\\
    X_{4,n} &= aV(1) - X_{4,t-1}^2 + aV(2)X_{4,t-2} - 0.02cX_{3,t-2} + w_{4,n}\\
    X_{5,t} &= aV(1) - (0.5c(X_{1,t-1}+X_{2,t-1}) + \\
               &  (1-c)X_{5,t-1})^2 + aV(2)X_{5,t-2} + w_{5,n}
    \label{eq:henonAle}
 \end{aligned}
\end{equation}
where \textit{aV} is the characteristic parameter tuned for chaotic behavior, the coupling strength $c = 0.4$ and \textit{w} is drawn from Gaussian noise with zero mean and unit variance. In figure \ref{simsysHenonAle} the modeled links are shown.

We performed the analysis setting $l_X = l_Y = l_Z = 5$ and using the rest of the parameters values found in Subsection \ref{correct_params}. Looking at sensitivity, specificity and $\mbox{F1}_{\mbox{score}}$, table \ref{tabdiscardue}, we can clearly notice that NeuNet NUE performs better than NeuNet UE. Considering this result, the heavy computational complexity and the lack of information about the only past states that can give information to the target concerning the use of NeuNet UE, led us to only take into account NeuNet NUE for further investigations.

\begin{table}[ht]
	\centering
		\begin{tabular}{cccc}
		   \hline
			& Sens & Spec & $\mbox{F1}_{\mbox{score}}$\\
		   \hline
			\textbf{NeuNet NUE} & 0.86 & 0.80 & 0.80\\
			\hline
			\textbf{NeuNet UE} & 0.69 & 0.93 & 0.77\\
			\hline
		\end{tabular}
		\caption{Sensitivity, specificity and $\mbox{F1}_{\mbox{score}}$ values obtained on the system \eqref{eq:henonAle} by NeuNet NUE and NeuNet UE.}
		\label{tabdiscardue}
	\end{table}
 
 \subsection{Simulated data: H\'enon maps}
 \label{simHM}
 
\begin{figure}[ht!]
\centering
    \includegraphics[width=0.7\textwidth]{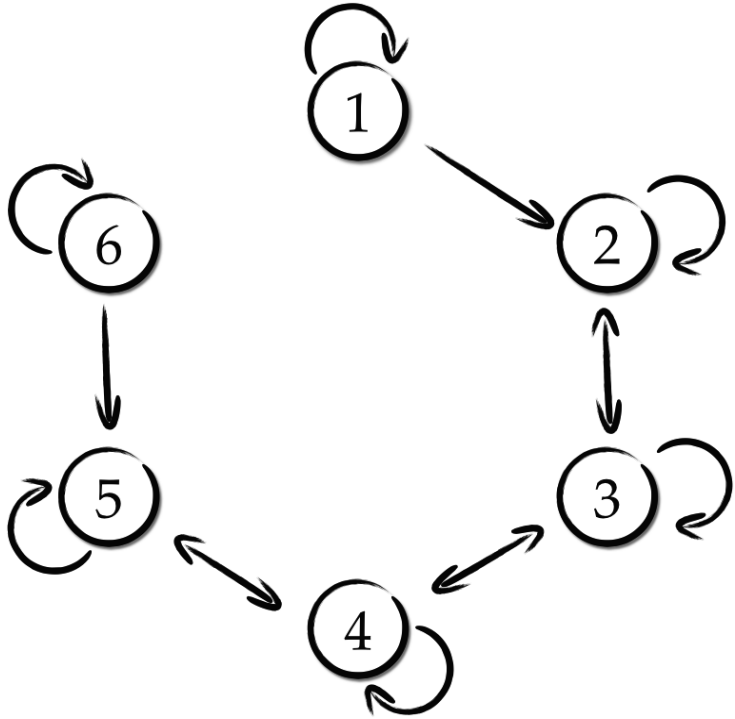}
    \caption{Simulated system. Interactions between the variables of the simulated H\'enon maps system generated according to equations \eqref{eq:henonmaps}.}
    \label{simsystHenon}
\end{figure}

 In the first experiment we generated 6 H\'enon maps rearranging the system \eqref{eq:henonAle} as shown in figure \ref{simsystHenon} setting the coupling strength $c = 0.2$. The equations are shown in the following

%
\begin{equation}
	\begin{aligned}
		X_{1,n}    &= aV(1) - X_{1,n-1}^2 + aV(2)X_{1,n-2} + w_{1,n} \\
		X_{m,n}   &= aV(1) - (0.5c_m(X_{m-1,n-1}+X_{m+1,n-1}) + \\
		             & + (1-c_m)X_{m,n-1})^2 + aV(2)X_{m,n-2} + w_{m,n} \\
		X_{6,n} &= aV(1) - X_{6,n-1}^2 + aV(2)X_{6,n-2} + w_{6,n}
		\label{eq:henonmaps}
	\end{aligned}
\end{equation}
where \textit{aV} is the vector of parameters for chaos, \textit{w} is drawn from Gaussian noise with zero mean and unit variance and $m \in [2,5]$ is the identifier of the series. 

\begin{figure}[ht!]
\centering
   \includegraphics[width=0.8\textwidth]{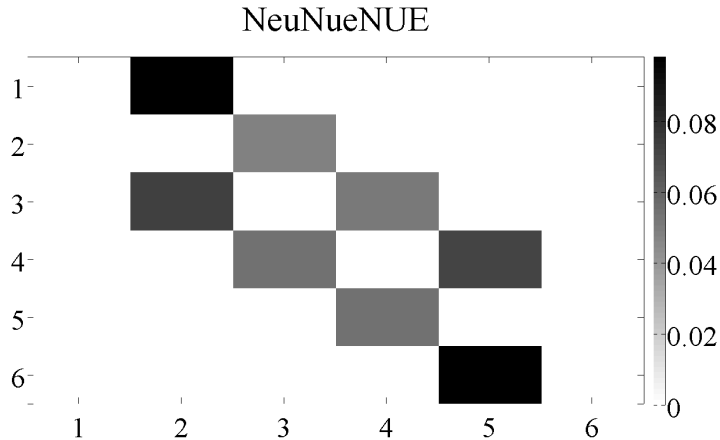}
    \caption{GC matrix representation for the NueNet NUE estimator applied to the system \eqref{eq:henonmaps}. The color indicates the magnitude of the GC averaged over 100 realizations of the simulation. The targets are plotted on the x-axis while on the drivers are plotted on the y-axis.}
        \label{neunet_henon}
\end{figure}

We generated 100 realizations of the H\'enon maps, performed a multivariate analysis keeping the parameters found in section \ref{correct_params} fixed and setting $l_X = l_Y = l_Z = 5$. We then evaluated the mean values of the NNGC for all the pairwise combinations driver-target as shown in figure \ref{neunet_henon}. 
We compared our method's performance with the binning, linear and nearest neighbor estimators implemented in the non-uniform embedding framework (henceforth BIN NUE, LIN NUE and NN NUE).
These three estimators are already implemented in MuTE \cite{montalto2014mute}. The comparison with NeuNet NUE has been performed in terms of sensitivity, specificity and F1$_{\mbox{score}}$, as shown in table \ref{tab2}. We can notice that NeuNet NUE is the second best method after NN NUE.

\begin{table}[ht]
	\centering
		\begin{tabular}{cccc}
		   \hline
			& Sens & Spec & $\mbox{F1}_{\mbox{score}}$\\
		   \hline
		   \textbf{BIN NUE} & 1 & 0.86 & 0.84\\
		   \hline
			\textbf{LIN NUE} & 0.99 & 0.74 & 0.73\\
			\hline
			\textbf{NeuNet NUE} & 1 & 0.98 & 0.98\\
			\hline
			\textbf{NN NUE} & 1 & 1 & 1\\
			\hline
		\end{tabular}
		\caption{Sensitivity, specificity and $\mbox{F1}_{\mbox{score}}$ values obtained on the system \eqref{eq:henonmaps} by the four estimators.}
		\label{tab2}
	\end{table}
	
\begin{figure}[ht!]
\centering
   \includegraphics[width=1\textwidth]{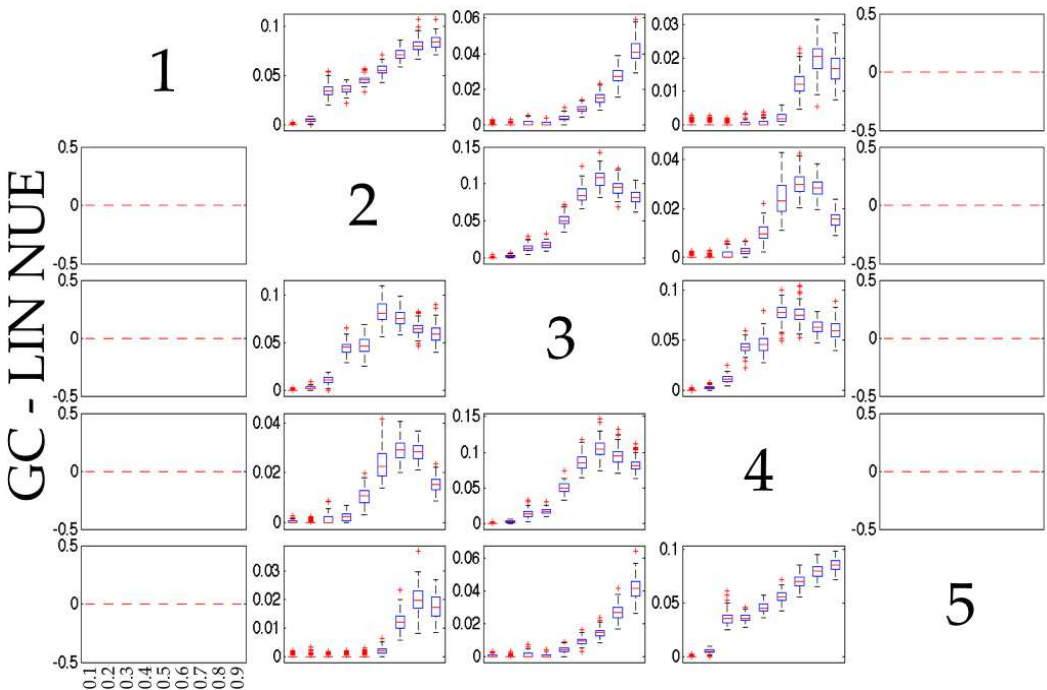}
    \caption{LIN NUE performances on H\'enon maps at varying of the coupling strength. GC values are plotted on the y-axis, while the coupling strength values are plotted on the x-axis.}
        \label{lin_henon_boxplot1}
\end{figure}

\begin{figure}[ht!]
\centering
   \includegraphics[width=1\textwidth]{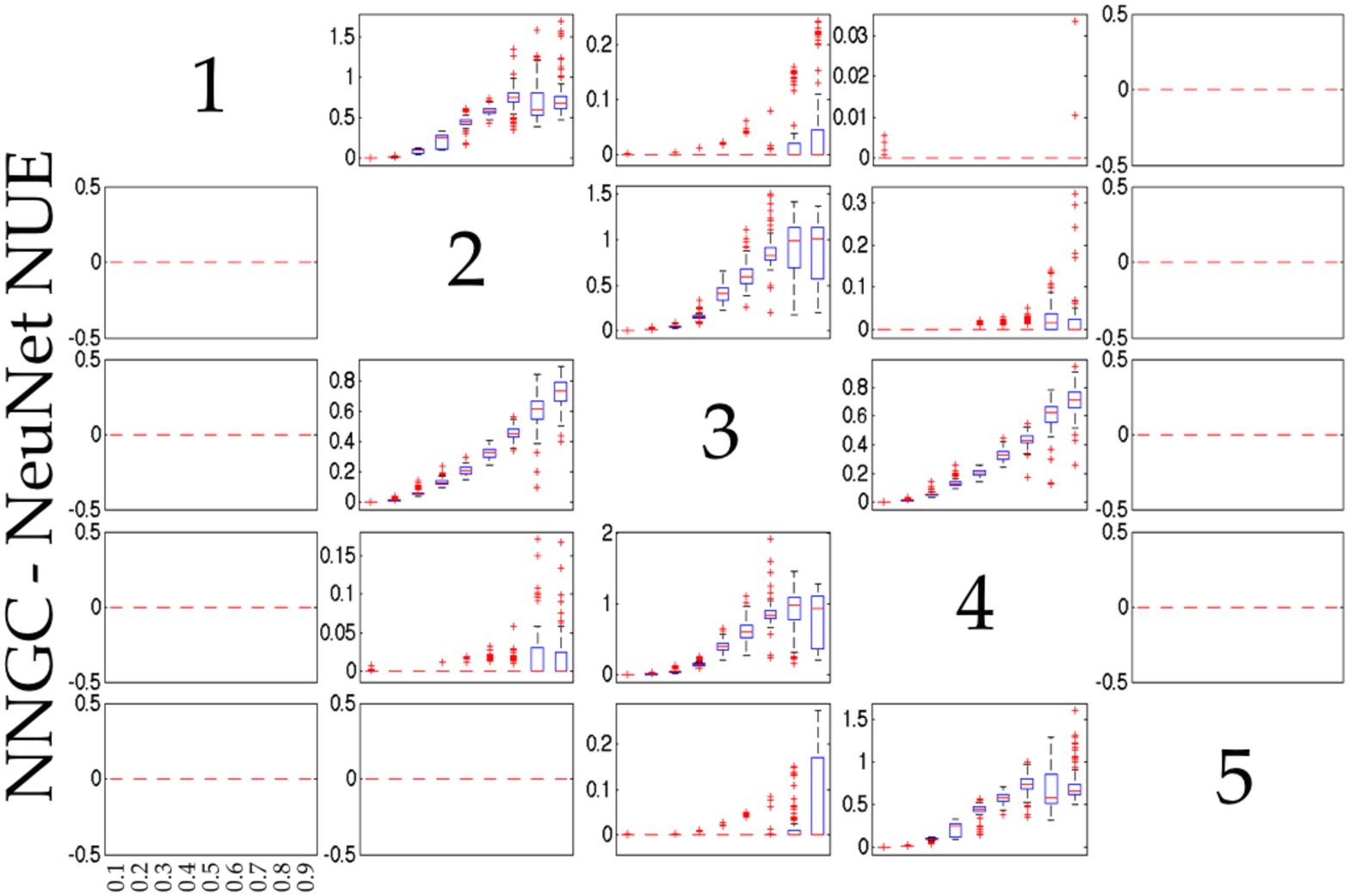}
    \caption{NeuNet NUE performances on H\'enon maps at varying of the coupling strength. NNGC values are plotted on the y-axis, while the coupling strength values are plotted on the x-axis.}
        \label{neunet_henon_boxplot1}
\end{figure}

\begin{figure}[ht!]
\centering
   \includegraphics[width=1\textwidth]{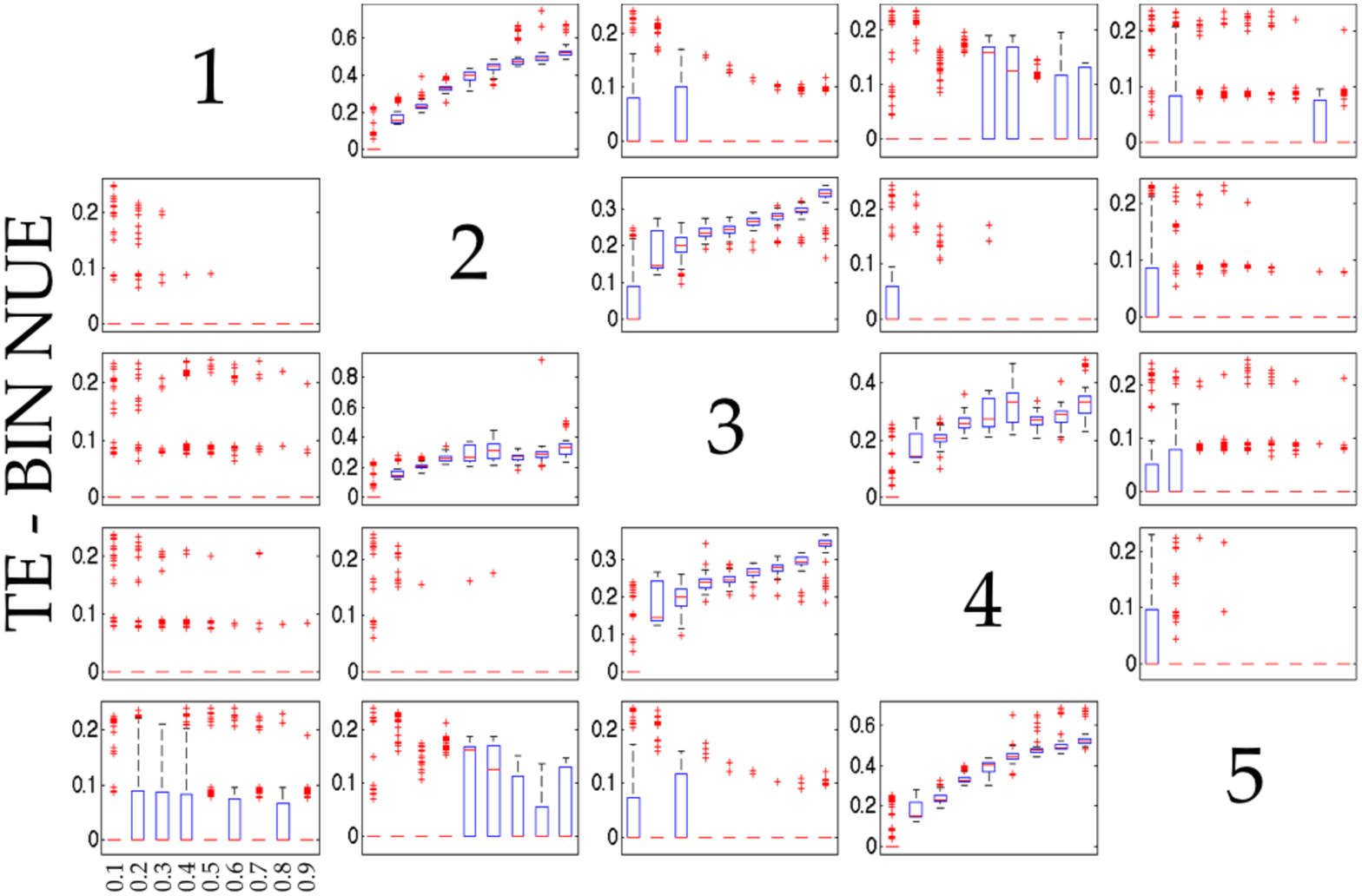}
    \caption{BIN NUE performances on H\'enon maps at  varying of the coupling strength. TE values are plotted on the y-axis, while the coupling strength values are plotted on the x-axis.}
        \label{bin_henon_boxplot2}
\end{figure}

\begin{figure}[ht!]
\centering
   \includegraphics[width=1\textwidth]{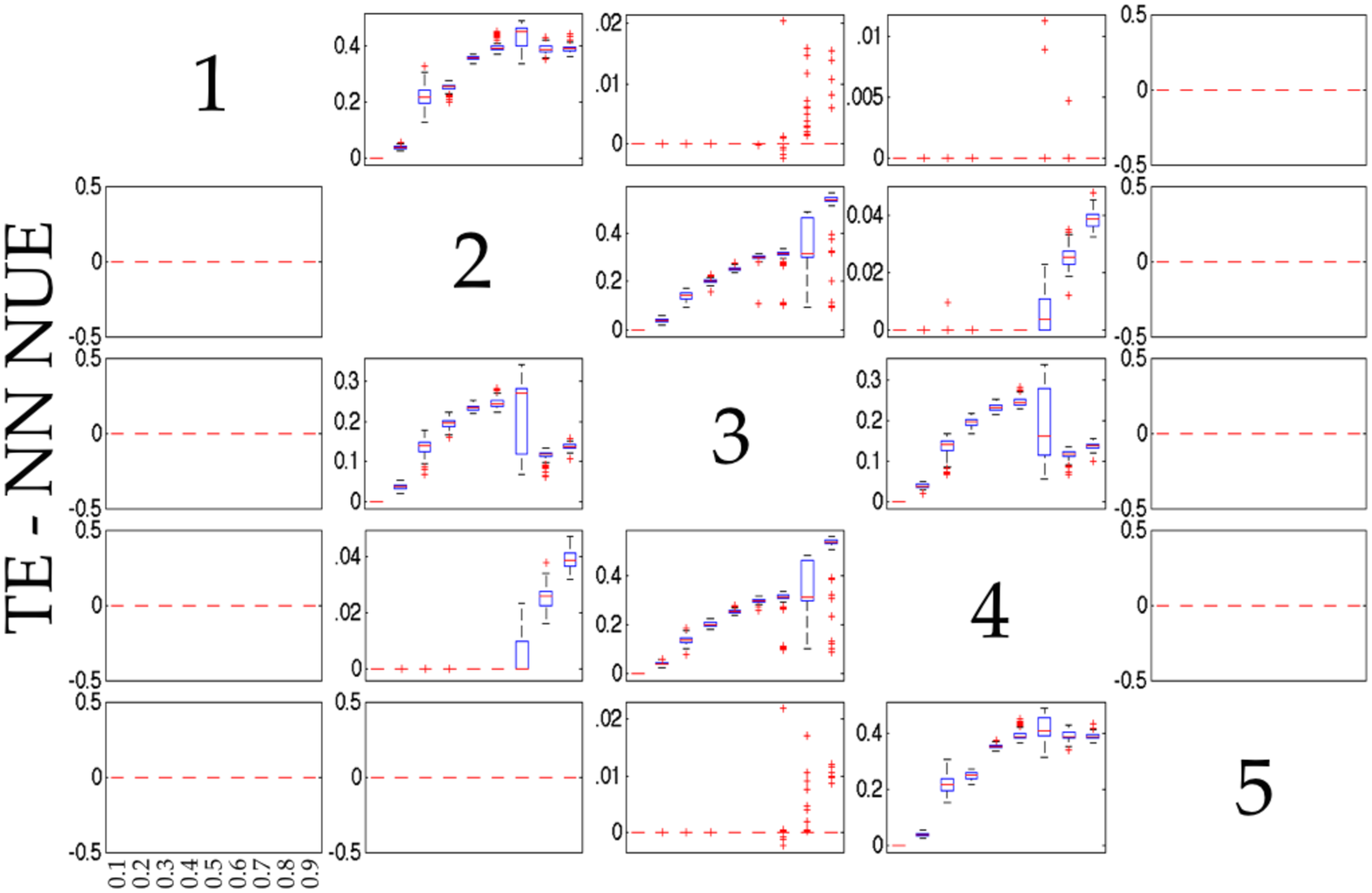}
    \caption{NN NUE performances on H\'enon maps at varying of the coupling strength. TE values are plotted on the y-axis, while the coupling strength values are plotted on the x-axis.}
        \label{nn_henon_boxplot3}
\end{figure}

Furthermore, we wanted to investigate whether NeuNet NUE was robust enough with respect to the coupling strength involved in \eqref{eq:henonmaps} using only 5 time series. Again we performed a comparison with the estimators implemented in MuTE. In figures \ref{lin_henon_boxplot1} - \ref{nn_henon_boxplot3} we can see the performances of the four methods, noticing that NeuNet NUE is the only approach able to detect the expected information flows even when the coupling is 0.8. NN NUE detects two false positive information flows for the directions $4 \rightarrow 2$, $2 \rightarrow 4$ from coupling strength value equal to 0.6 on. BIN NUE and LIN NUE obtain the worst performances detecting false positive information flows even for coupling strength equal to 0.3, see figure \ref{lin_henon_boxplot1} direction $2 \rightarrow 4$. Note the differing number of outliers in figure \ref{lin_henon_boxplot1} versus figure \ref{nn_henon_boxplot3}, even if the detection criterion is fixed: on each box, the central mark is the median, the edges of the box are the 25th and 75th percentiles, the whiskers extend to the most extreme data points not considered outliers.
The four methods show different fluctuations of the exchanged information values as remarked by the different number of outliers.

\begin{figure}[ht!]
\centering
   \includegraphics[width=1\textwidth]{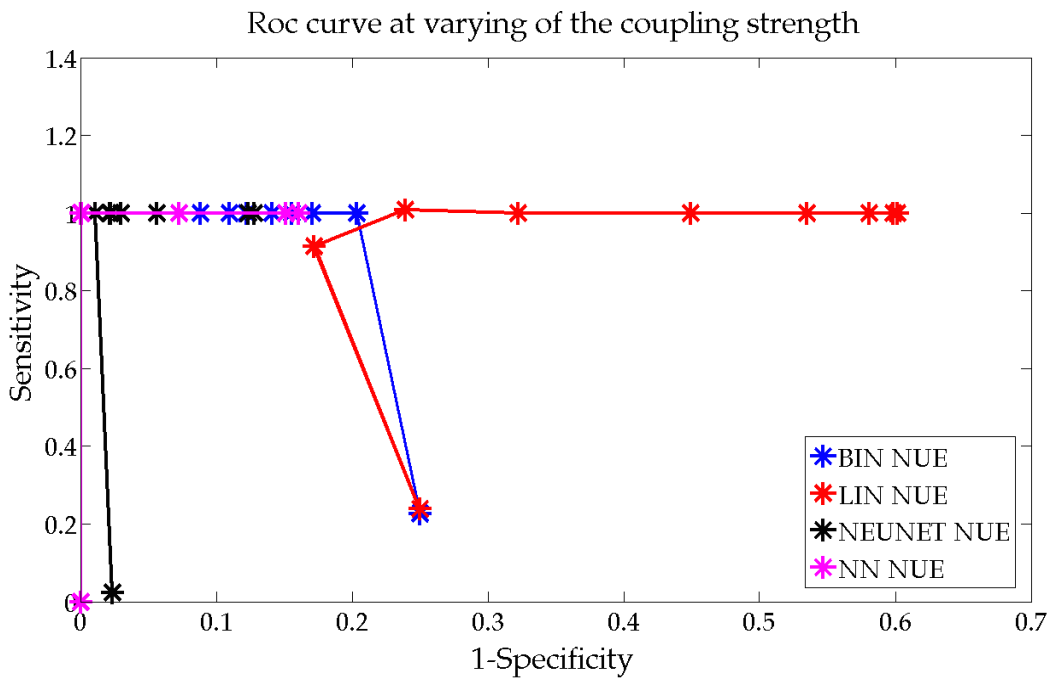}
    \caption{ROC curves obtained on H\'enon maps at varying of the coupling strength.}
        \label{roc_curve_coupling}
\end{figure}

\begin{figure}[ht!]
\centering
   \includegraphics[width=1\textwidth]{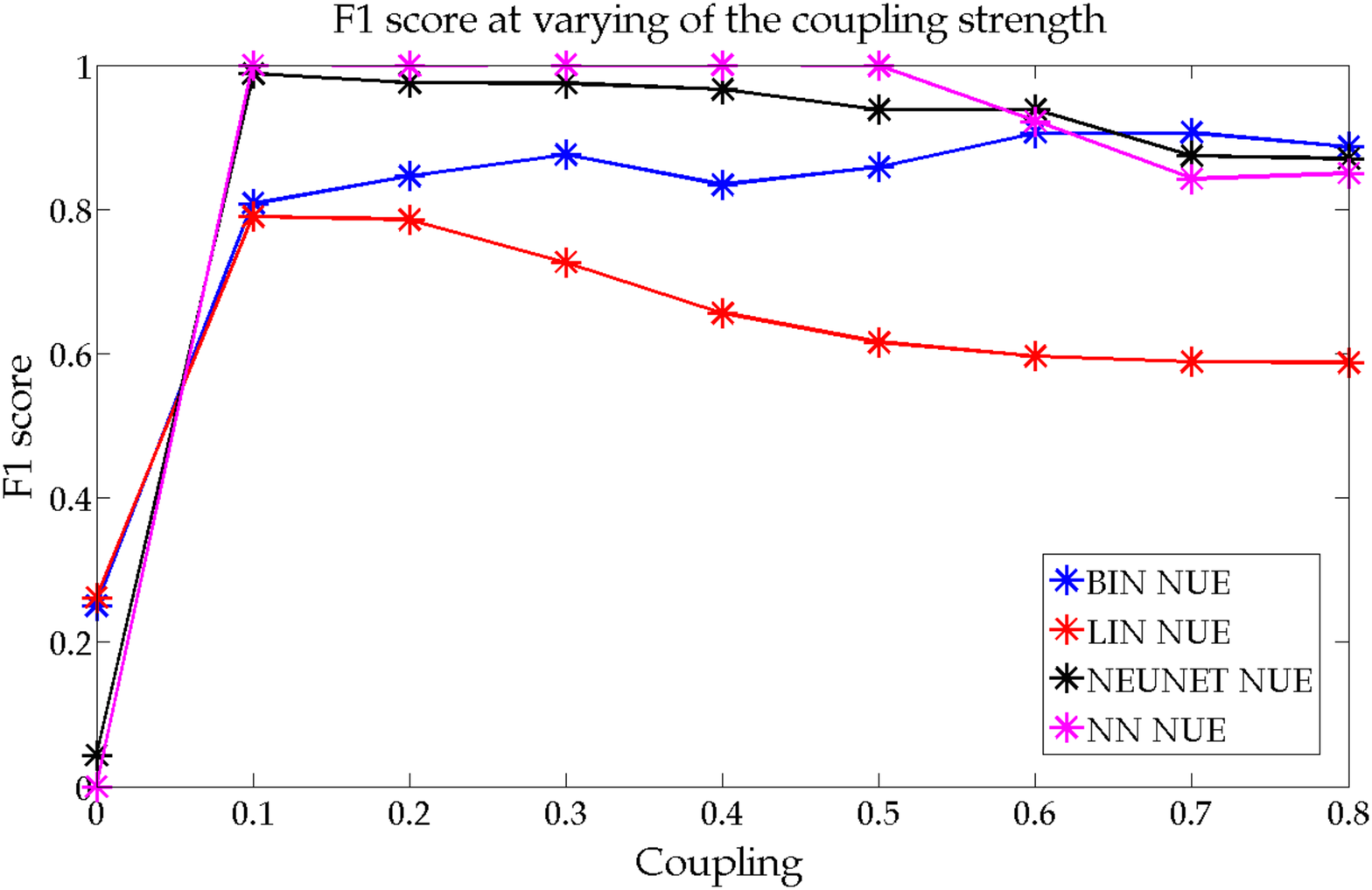}
    \caption{$\mbox{F1}_{\mbox{score}}$ obtained on H\'enon maps at varying of the coupling strength.}
        \label{f1score_coupling}
\end{figure}

In figures \ref{roc_curve_coupling}, \ref{f1score_coupling} we compare the performances of the four methods showing their ROC curves and $\mbox{F1}_{\mbox{score}}$, respectively. NN NUE and NeuNet NUE ROC curves report the highest sensitivity and specificity as soon as the coupling is greater than zero. For high couplings the ROC curves denote a higher specificity of NeuNet NUE. BIN NUE starts with low sensitivity and specificity, and its specificity generally increases as the coupling increases. $\mbox{F1}_{\mbox{score}}$ curves belonging to NeuNet NUE and NN NUE are very close. For couplings greater than 0.5 NeuNet NUE $\mbox{F1}_{\mbox{score}}$ is higher than NN NUE $\mbox{F1}_{\mbox{score}}$, but both lower than BIN NUE $\mbox{F1}_{\mbox{score}}$ denoting once again how NeuNet NUE can be much closer to model-free than to model-based approaches. Only at coupling = 0.6 NeuNet NUE has the highest $\mbox{F1}_{\mbox{score}}$.

Referring to the equations \eqref{eq:henonAle}, we obtained the results shown in table \ref{tab:henonAle} in terms of sensitivity, specificity and F1$_{\mbox{score}}$. This time NeuNet NUE detects causal influences better than BIN NUE, even if it is still not able to perform better than NN NUE. 

\begin{table}[ht]
	\centering
		\begin{tabular}{cccc}
		   \hline
			& Sens & Spec & $\mbox{F1}_{\mbox{score}}$\\
		   \hline
		   \textbf{BIN NUE} & 0.85 & 0.68 & 0.73\\
		   \hline
			\textbf{LIN NUE} & 0.88 & 0.45 & 0.65\\
			\hline
			\textbf{NeuNet NUE} & 0.86 & 0.80 & 0.80\\
			\hline
			\textbf{NN NUE} & 0.87 & 0.91 & 87\\
			\hline
		\end{tabular}
		\caption{Sensitivity, specificity and $\mbox{F1}_{\mbox{score}}$ values obtained on the system \eqref{eq:henonAle} by the four estimators}
		\label{tab:henonAle}
	\end{table}

\subsection{Simulated data: Lorenz system}
\label{simLS}

In the third experiment we studied a system composed of five identical Lorenz subsystems defined by the following equations:

\begin{equation}
  \begin{aligned}
     \dot{x}_1 &= -10x_1 +10x_1, \;\;\; & \dot{x}_i &= -10x_i +10x_i +C(x_{i-1} - x_i),  \\
     \dot{y}_1 &= -x_1z_1 + 28x_1 - y_1, \;\;\;& \dot{y}_i &= -x_iz_i + 28x_i - y_i,\\
     \dot{z}_1 &= x_1y_1 - 8/3z_1, \;\;\;&\dot{z}_i &= x_iy_i - 8/3z_i, \\
  \end{aligned}
  \label{eq:lorenz}
\end{equation}
where $i \in [2,5]$. The differential equations are solved by means of the Runge-Kutta method implemented in MATLAB and the time series are generated at a sampling rate of 0.01 time units. The subsystems, ranging from $X_1$ to $X_5$, influence each other according the following rule: \textit{i}-th time series is influenced only by the $(i-1)$-th time series except for $X_1$ that only gives influence to $X_2$. The coupling strength $C = 5$ is the same for the whole set on influences.

The nature of the Lorenz system results in a more challenging system than the H\'enon systems. The parameters set up have been kept the same as in the other experiments. Even in this scenario NeuNet NUE can reach good performances with both high sensitivity and specificity, as shown in table \ref{tab:lorenz}. Our method, on this system too, reaches performances in the middle between the model-free approaches and the model-based approach.

\begin{table}[ht]
	\centering
		\begin{tabular}{cccc}
		   \hline
			& Sens & Spec & $\mbox{F1}_{\mbox{score}}$\\
		   \hline
		   \textbf{BIN NUE} & 0.94 & 0.91 & 0.82\\
		   \hline
			\textbf{LIN NUE} & 0.51 & 0.72 & 0.39\\
			\hline
			\textbf{NeuNet NUE} & 0.70 & 0.95 & 0.74\\
			\hline
			\textbf{NN NUE} & 1 & 0.86 & 0.77\\
			\hline
		\end{tabular}
		\caption{Sensitivity, specificity and $\mbox{F1}_{\mbox{score}}$ values obtained on the Lorenz system by the four estimators.}
		\label{tab:lorenz}
	\end{table}
	
\begin{figure}[ht!]
\centering
   \includegraphics[width=1\textwidth]{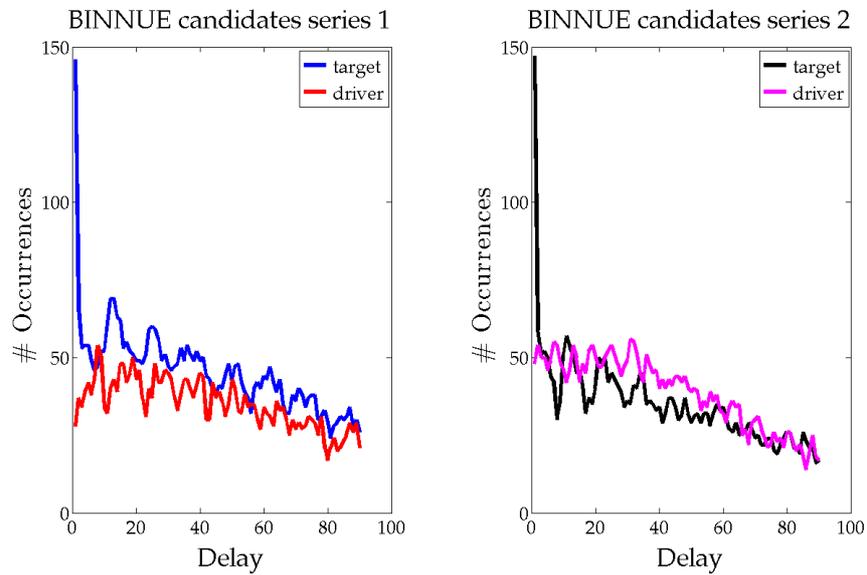}
    \caption{BIN NUE performances on bidirectionally coupled Lorents system. }
        \label{binnue_candidates}
\end{figure}

\begin{figure}[ht!]
\centering
   \includegraphics[width=1\textwidth]{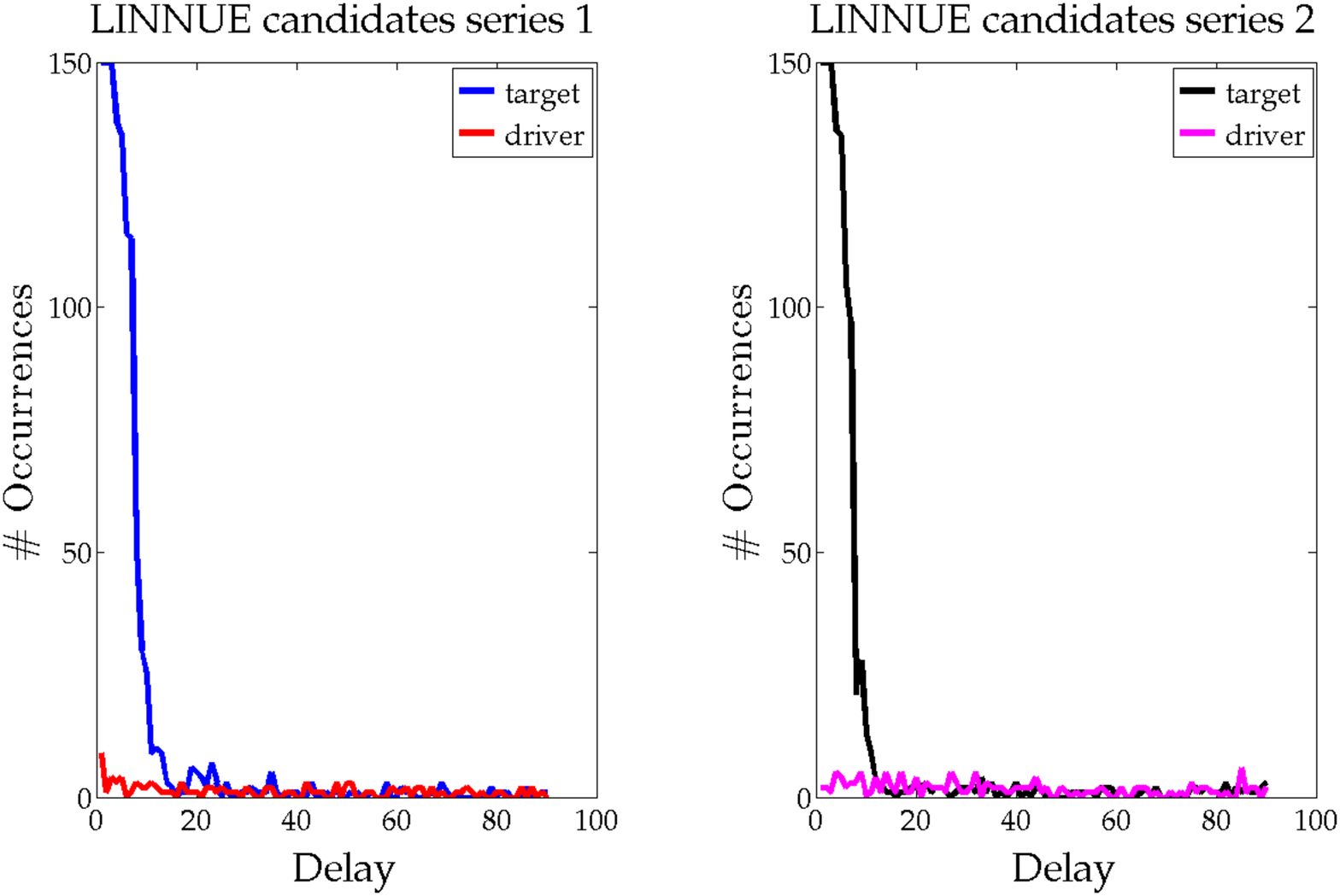}
    \caption{LIN NUE performances on bidirectionally coupled Lorents system. }
        \label{linnue_candidates}
\end{figure}

\begin{figure}[ht!]
\centering
   \includegraphics[width=1\textwidth]{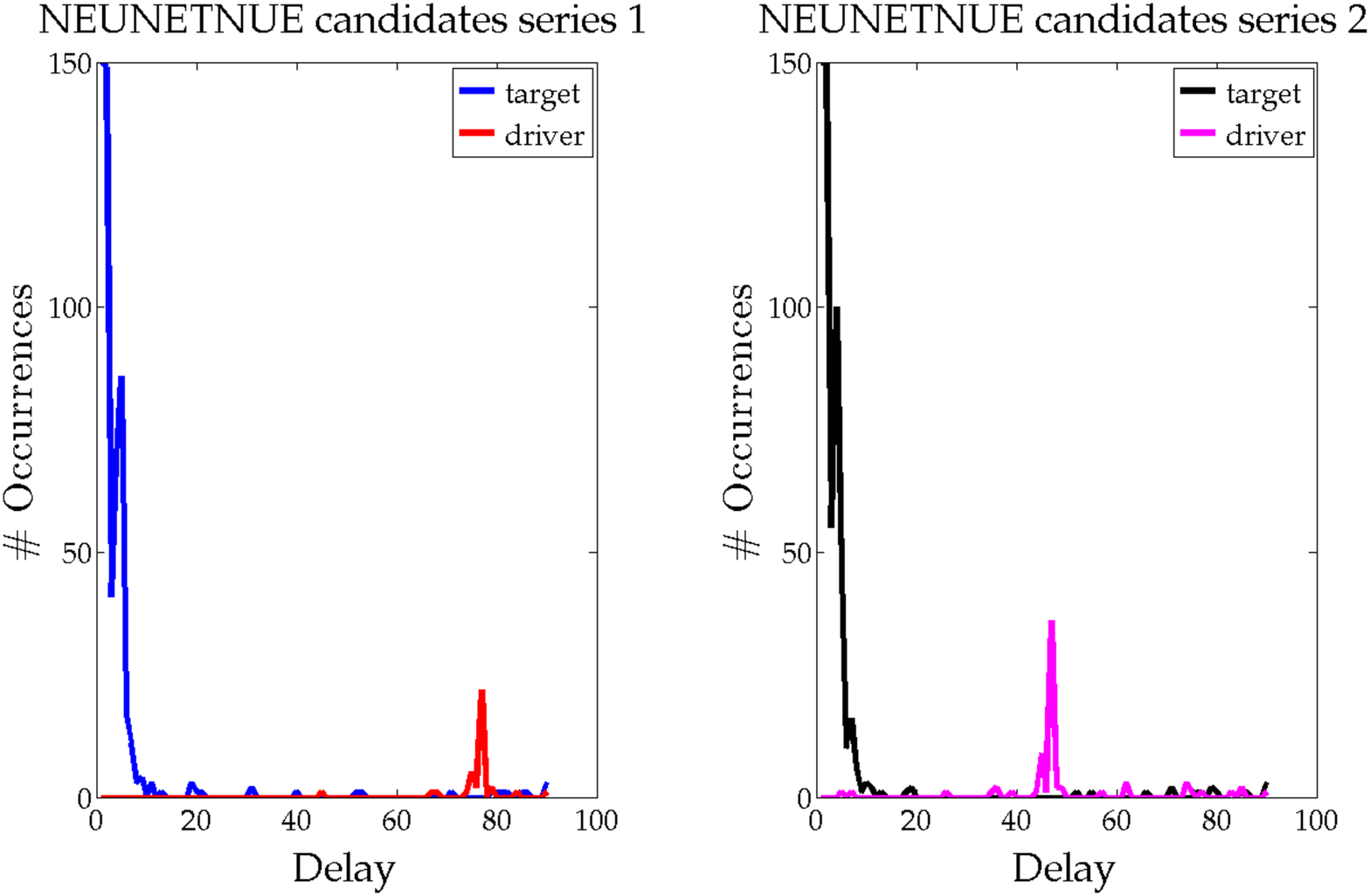}
    \caption{NeuNet NUE performances on bidirectionally coupled Lorents system. }
        \label{neunetnue_candidates}
\end{figure}

\begin{figure}[ht!]
\centering
   \includegraphics[width=1\textwidth]{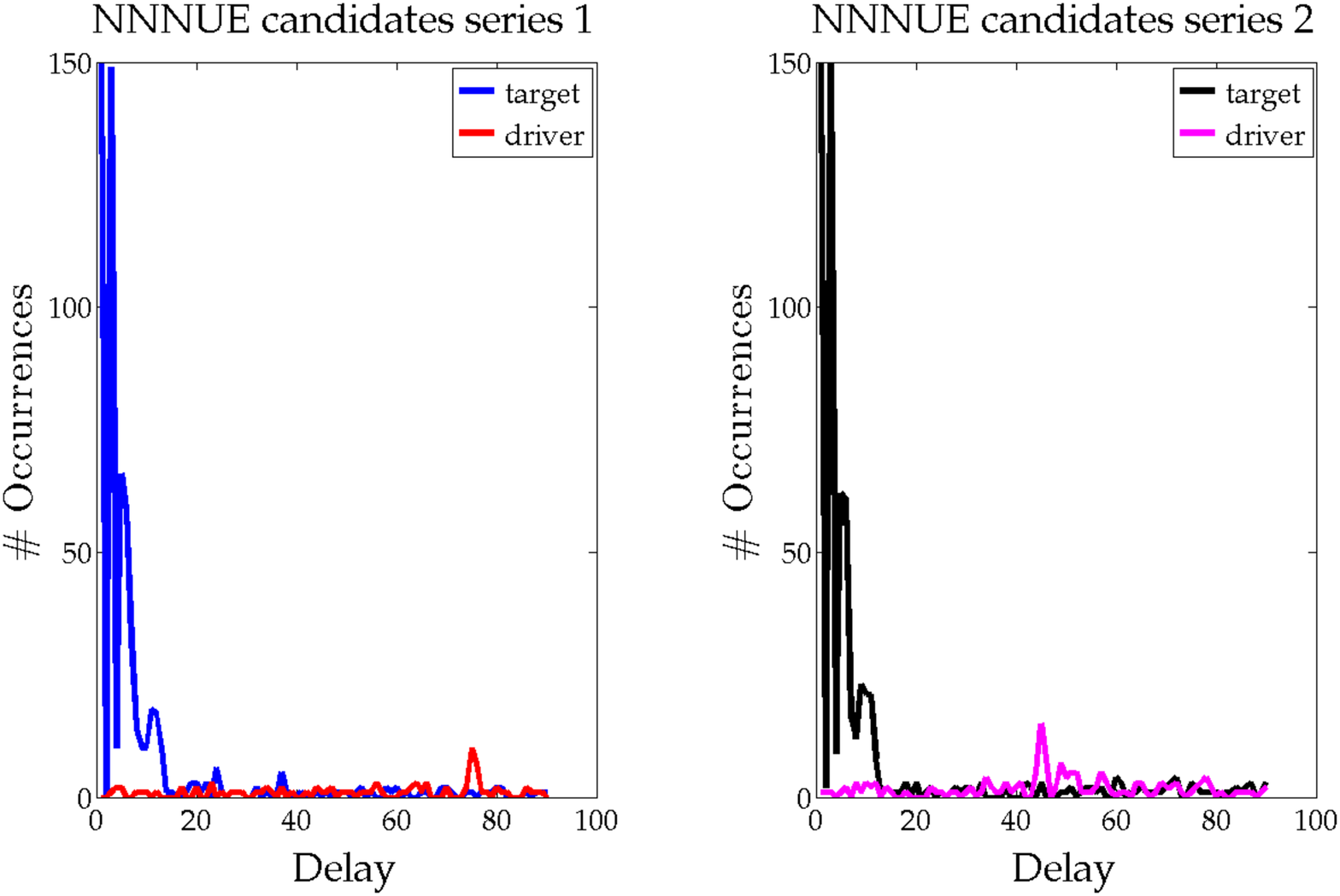}
    \caption{NN NUE performances on bidirectionally coupled Lorents system. }
        \label{nnnue_candidates}
\end{figure}

Another experiment on a chaotic Lorenz system was performed in order to check how robust NeuNEt NUE could be with respect to influences occurring at longer delays. We used 150 bidirectionally coupled Lorenz systems as  in \cite{wibral2013measuring}. The delay at which series 1 influences series 2 was set at 45 points back, while the delay at which series 2 influences series 1 was set at 75 points back. The coupling constant was set as 0.1 for both series. We chose 90 candidates for each series and checked how many times each candidate was chosen. As we can see in figures \ref{binnue_candidates}-\ref{nnnue_candidates} NN NUE can detect the right delays even if there are many other candidates chosen. NeuNet NUE was successful in retrieving the correct influences at the corresponding delays, more often than NN NUE as it can be seen from the height of the peaks. The other two estimators clearly failed in detecting the right influences and delays.

\subsection{Redundant data}
\label{simRD}

An issue that complicates the correct detection of GC is the presence of redundant variables. In this case the conditioning approach is misled and the analysis results in false negatives (see \cite{stramaglia2014synergy} for a complete explanation of this phenomenon).
We applied neural networks Granger causality analysis to redundant data to check whether the approach was able to detect the right information flows with an increasing number of redundant variables. We used data generated by the following equations:

\begin{equation}
  \begin{aligned}
  t_n &= h_{n-2} + c\varepsilon_n\\
  d_{i,n} &= h_{n-1} +c\varphi_n
  \label{eq:redundant}
  \end{aligned}
\end{equation}
where the process \textit{h} and the noises $\varepsilon, \varphi$ are drawn from a Gaussian distribution with zero mean and unit variance. The coefficient \textit{c} modulates the noise. The system represent a chain of influences, for which redundancy arises when $i \geq 3$, (the first two variables share information on the future of the third one), and so on.

\begin{figure}[ht!]
\centering
   \includegraphics[width=0.8\textwidth]{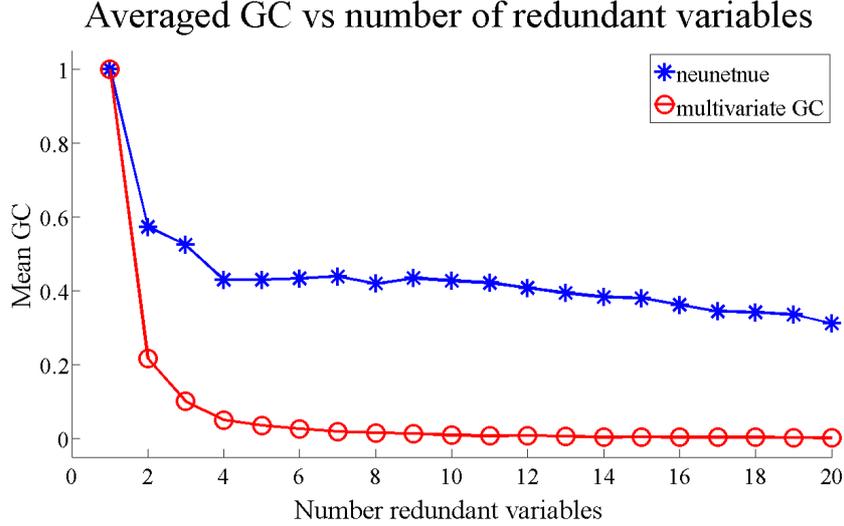}
    \caption{NeuNet NUE and multivariate GC performances on the redundant system.}
        \label{neunet_mvgc_redundant}
\end{figure}

We compared NeuNet NUE with the fully conditioned non-linear kernel Granger causality as in \cite{stramaglia2014synergy}. The experiments were performed with $l_X = l_Y = l_Z = 5$ and keeping the parameters found in Subsection \ref{correct_params} fixed. We generated 20 trials of the system \eqref{eq:redundant} varying the number of redundant variables from 1 up to 20, with 2500 time points. The analyses were performed taking into account the variable \textit{t} as targets and each variable $d_i$ as driver, conditioning on the remaining $d_{(i-1)}$ variables, with $i \in [1,\ldots,20]$. We then evaluated GC for both methods averaging over the number of trials, varying the number of redundant variables. According to the results shown in figure \ref{neunet_mvgc_redundant}, we can notice how GC detected by the neural networks never drops to zero, as it happens for kernel GC. Table \ref{tab:redundancy} reports the number of false negatives given by NeuNet NUE. It is worth noting that the amount of false negatives is zero up to 10 redundant variables. Conversely, due to the different construction of the method, the values of kernel Granger causality are always significant, albeit very low, at least for this system size.

\begin{table}[ht]
	\centering
		\begin{tabular}{p{1.2cm}p{0.08cm}p{0.08cm}p{0.08cm}p{0.08cm}p{0.08cm}p{0.08cm}p{0.08cm}p{0.08cm}p{0.08cm}p{0.08cm}p{0.08cm}p{0.08cm}p{0.08cm}p{0.08cm}p{0.08cm}p{0.08cm}p{0.08cm}p{0.08cm}p{0.08cm}p{0.08cm}p{0.08cm}}
		   \hline
			\# RV & 1 & 2 & 3 & 4 & 5 & 6 & 7 & 8 & 9 & 10 & 11 & 12 & 13 & 14 & 15 & 16 & 17 & 18 & 19 & 20\\
		   \hline
		   \# FN & 0 & 0 & 0  & 0 & 0 & 0 & 0 & 0 & 0 & 0 & 3 & 4 & 14 & 13 & 17 & 29 & 41 & 49 & 57 & 79\\
			\hline
		\end{tabular}
		\caption{Number of false negatives, FN, returned by NeuNet NUE for 20 trials at varying of the number of redundant variables, RV.}
		\label{tab:redundancy}
	\end{table}

\subsection{Classification task}
\label{CT}

Our final goal in this paper was to test whether NeuNet NUE could correctly classify dynamics. We trained NeuNet NUE on the system \eqref{eq:henonmaps}. We have randomly chosen one of the networks trained to detect the causal influences towards a certain target. Then we fed the network with 100 realizations of the system \eqref{eq:henonmaps}, never used in the learning phase, and with 100 realizations of an autoregressive system, represented by the following equations:

\begin{equation}
\begin{aligned}
	X_{1,n} &=  0.95 \sqrt{2} \, X_{1,n-1} - 0.9025\, X_{1,n-2} + z_{1,n} \\
	X_{2,n} &= 0.5\, X_{1,n-2}^2 + z_{2,n}  \\
	X_{3,n} &= -0.4\, X_{1,n-3} + z_{3,n}  \\
	X_{4,n} &= -0.5\, X_{1,n-2}^2 + 0.25 \sqrt{2} \, X_{4,n-1} + 0.25 \sqrt{2} \, X_{5,n-1} + z_{4,n}  \\
	X_{5,n} &= -0.25 \sqrt{2} \, X_{4,n-1} + 0.25 \sqrt{2} \, X_{5,n-1} + z_{5,n}  
	\label{eq:nonlinearmodel}
\end{aligned}
\end{equation}
where $z_{1,n}, z_{2,n}, z_{3,n}, z_{4,n}, z_{5,n}$ are drawn from Gaussian noise with zero mean and unit variance.

\begin{figure}[ht!]
\centering
   \includegraphics[width=0.8\textwidth]{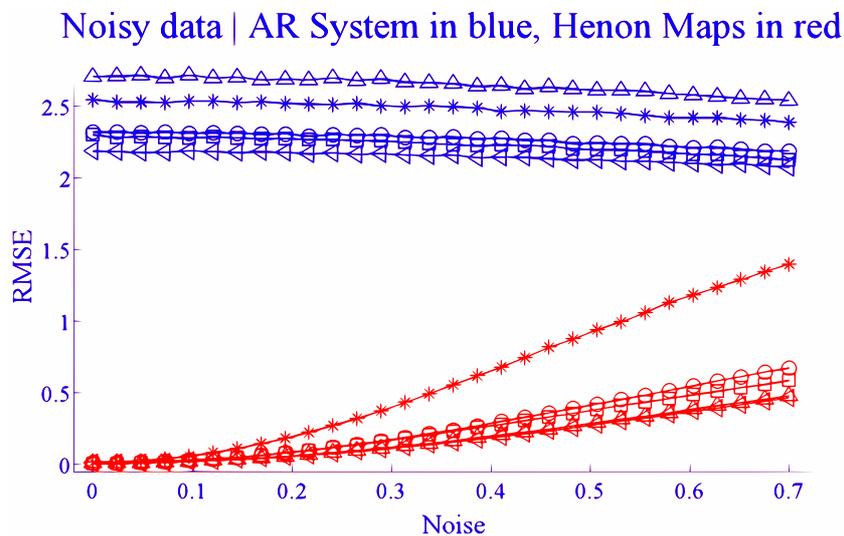}
    \caption{RMSE versus the noise level. The red curves are obtained testing NeuNet NUE on the same kind of data with which it was trained. The blue curves are obtained testing NeuNet NUE on the system \eqref{eq:nonlinearmodel}. Each curve represents the network trained to detect the influence towards a specific target with a different coupling strength.}
        \label{neunet_classification}
\end{figure}

System \ref{eq:henonmaps} is considered, this time with only five variables to be consistent with the size of the autoregressive model. We then evaluated the average RMSE for both the systems. We repeated the procedure for 30 different noise values, ranging from 0 to 0.7, and different couplings strength, ranging in the interval [0,0.8] with step of 0.2. The results are shown in figure \ref{neunet_classification}. We plotted the average RMSE with respect to the different noise level for each coupling strength value. We can notice that the errors obtained when the two systems are given to NeuNet NUE as test sets lie in linear separable portions of space. This represents an encouraging result as it may be useful to classify systems, given that our approach has been trained with a known system.

\section{Conclusions}

In this paper we have implemented the Granger paradigm for detection of dynamical influences in the frame of feed-forward neural networks. The novelty of the present approach arises from the use of non-uniform embedding for variable selection and generalization error for the assessment of Granger causality. We have demonstrated the theoretical and experimental advantages of implementing the neural network approach with non-uniform embedding compared to the uniform one. Due to the universal character of function approximation of neural networks, the proposed approach is intermediate between the classical Granger linear implementation and the non-parametric estimator corresponding to transfer entropy: by means of several examples, we have shown that there are situations where our approach outperforms both approaches. The proposed method differs from the kernel Granger causality not only by providing a validation phase, but also by letting the neural networks explore the parameters space and building the best model to explain the information transfers among variables. Kernel Granger causality, instead, is still a model-based approach, for which the type of kernel and the degree of non-linearity have to be specified beforehand. We would like to remark that so far neural networks have been used to detect GC only when combined with other estimators. Furthermore the training phase was stopped only when a certain number or training epochs was reached. This choice seems quite approximate because of the lack of knowledge about the exact amount of training epochs needed to both minimize the error function and to avoid overfitting the neural networks. Therefore, the validation phase is necessary in our opinion to ensure that the network fully explores the parameters space, converges to a minimum and avoids the risk of overfitting.
We conclude remarking that other wrappers can be taken into account and many deep learning architectures are built from artificial neural networks, therefore we expect that  further developments of our approach will be the implementations of Granger causality both using other feature selection algorithms and in the frame of deep learning \cite{SIG-039}.

\section*{Acknowledgments}

This work is supported by: the Belgian Science Policy (IUAP VII
project CEREBNET P7 11); the University of Ghent (Special Research Funds for visiting researchers).
The neural network toolbox has been developed by Roberto Prevete and Giovanni Tessitore:
rprevete@unina.it;
tessitore@na.infn.it

The authors would like to thank Dr. Christopher Stewart for his valuable feedback.

\bibliographystyle{unsrt}
\bibliography{articleNeuNet_bib}

\end{document}